%% file: main.tex
\documentclass[11pt]{article}

\usepackage[left=1in,right=1in,top=1in,bottom=1in]{geometry}
\input{math_commands.tex}

\usepackage{amssymb, amsthm, mathtools, booktabs, graphicx, microtype, xcolor}
\usepackage{subcaption}
\usepackage{upgreek}   %
\usepackage{nomencl}   %
\usepackage[ruled]{algorithm2e}

\SetAlFnt{\normalsize}
\SetAlCapFnt{\normalsize}
\SetAlCapNameFnt{\normalsize}
\SetAlCapHSkip{0pt}
\IncMargin{-\parindent}
\usepackage[square,numbers]{natbib}
\usepackage[hidelinks,pageanchor]{hyperref}
\usepackage{aliascnt} %
\usepackage[capitalise]{cleveref}  %

\definecolor{darkpastelgreen}{rgb}{0.01, 0.75, 0.24}
\usepackage{nomencl}

\newcommand{\newsharedtheorem}[3]{%
  \newaliascnt{#1}{theorem}%
  \newtheorem{#1}[#1]{#2}%
  \aliascntresetthe{#1}%
  \crefname{#1}{#2}{#3}%
  \Crefname{#1}{#2}{#3}%
}
\theoremstyle{plain}
\newtheorem{theorem}{Theorem}[section]
\newsharedtheorem{lemma}{Lemma}{Lemmas}
\newsharedtheorem{proposition}{Proposition}{Propositions}
\newsharedtheorem{corollary}{Corollary}{Corollaries}
\newsharedtheorem{claim}{Claim}{Claims}
\newsharedtheorem{conjecture}{Conjecture}{Conjectures}

\theoremstyle{definition}
\newsharedtheorem{definition}{Definition}{Definitions}
\newsharedtheorem{assumption}{Assumption}{Assumptions}
\newsharedtheorem{example}{Example}{Examples}

\theoremstyle{remark}
\newsharedtheorem{remark}{Remark}{Remarks}

\crefname{assumption}{Assumption}{Assumptions}
\Crefname{assumption}{Assumption}{Assumptions}

\newcommand{\veta}{\boldsymbol{\eta}}

\newcommand{\vtheta}{\boldsymbol{\theta}}
\newcommand{\valpha}{\boldsymbol{\alpha}}

\newcommand{\BigParens}[1]{\Bigl(#1\Bigr)}

\newcommand{\symm}{\mathfrak{S}}

\newcommand\defn[1]{\emph{\textbf{#1}}}

\title{Information Elicitation by Ordinal Agreement}
\author{
  Yiling Chen\thanks{Authors are listed in alphabetical order.}\\ Harvard University
  \and Shi Feng\\ Harvard University
  \and Grant Schoenebeck\\ University of Michigan
  \and Fang-Yi Yu\\ George Mason University
}
\date{}

\begin{document}
\maketitle

\begin{abstract}
\input{sections/abstract}
\end{abstract}

\input{sections/body}

\bibliographystyle{plainnat}
\bibliography{reference}

\clearpage
\appendix

\crefalias{section}{appendix}
\input{sections/appendix}
\end{document}

%% file: math_commands.tex
\usepackage{amsmath,amsfonts,bm}

\def\eqref#1{equation~\ref{#1}}

\def\1{\bm{1}}

\def\rx{{\textnormal{x}}}

\def\rva{{\mathbf{a}}}

\def\rvr{{\mathbf{r}}}

\def\rvx{{\mathbf{x}}}

\def\rmX{{\mathbf{X}}}

\def\vone{{\bm{1}}}
\def\va{{\bm{a}}}

\def\vc{{\bm{c}}}

\def\vr{{\bm{r}}}

\def\vx{{\bm{x}}}
\def\vy{{\bm{y}}}
\def\vz{{\bm{z}}}

\def\mA{{\bm{A}}}
\def\mB{{\bm{B}}}

\def\mD{{\bm{D}}}

\def\mP{{\bm{P}}}

\def\mX{{\bm{X}}}

\DeclareMathAlphabet{\mathsfit}{\encodingdefault}{\sfdefault}{m}{sl}
\SetMathAlphabet{\mathsfit}{bold}{\encodingdefault}{\sfdefault}{bx}{n}

\def\gS{{\mathcal{S}}}

\newcommand{\E}{\mathbb{E}}

\newcommand{\R}{\mathbb{R}}

\DeclareMathOperator{\sign}{sign}

%% file: sections/abstract.tex
Ordinal information is central to tasks ranging from conference peer review to Reinforcement Learning from Human Feedback for large language models. In many such settings, there is no objective ground truth against which reports can be directly verified, making incentive-compatible elicitation challenging. Peer-prediction mechanisms have been designed to achieve truthful elicitation despite the lack of ground-truth verification. However, existing peer-prediction methods do not provide incentives for agents to truthfully compress rich private information (e.g. reviews) into coarser ordinal reports (e.g. rankings) requested by the mechanisms.  
We propose a simple ordinal peer-prediction mechanism that rewards agents based on pairwise agreement, measured by Kendall tau. Under a broad class of Bayesian Stochastically Order Consistent models---which includes classical models such as Mallows and Thurstone---we prove that truthful reporting is stochastically dominant, and strongly truthful under index-equivariant strategies. This provides robust incentives even for agents with non-linear utility functions, such as risk aversion. Our mechanism further accommodates heterogeneous agents, top-K reports, and subset assignments, where each agent is scored against one randomly drawn peer. Experiments on synthetic Mallows and Thurstone data, together with a real-data stress test on the SUSHI preference dataset, show right-shifted truthful-payment distributions and the highest normalized sensitivity among tested baselines.

%% file: sections/body.tex
\section{Introduction}

Many important evaluation tasks require eliciting ordinal information. Conference program committees rank submissions for acceptance; platforms collect human feedback to compare the quality of LLM responses; and instructors use peer assessment to ask students to rank a small set of assignment submissions. These settings share a common structure: (1) Evaluators often have informative but imperfect signals to judge relative quality. (2) Investing more effort leads to higher-quality judgments. (3) There is often no objective ground truth against which reports can be directly verified.

This absence of ground truth creates a fundamental incentive problem. Without appropriate incentives, agents may invest little effort in evaluation or report strategically. Directly rewarding ``correct'' answers is not possible when correctness cannot be verified. The peer-prediction literature addresses the challenge of elicitation without ground truth by rewarding agents based on the statistical relationships among their reports. When properly designed, such mechanisms make truthful reporting an equilibrium even in the absence of verifiable outcomes. However, most existing peer-prediction mechanisms require agents to report all of their private information~\cite{prelec2004bayesian,shnayder2016informed,kong2019information}, typically as a numerical or categorical signal, and their incentive guarantees can fail when agents observe richer information than they are asked to report~\cite{frongillo2025binaryreportpeerpredictionrealvalued}. This creates a mismatch in many evaluation settings: agents may observe rich private information, such as detailed reviews, latent quality assessments, or noisy cardinal scores, while the designer only wants a coarser ordinal report, such as a ranking or a set of pairwise comparisons.

This paper studies peer prediction for coarse ordinal elicitation. Agents may observe rich private information about item quality, but the mechanism asks them to report only ordinal information. We allow several natural ordinal report formats, including full rankings, fixed-shape bucket orders such as top-$K$ sets and tiered reports, and rankings over assigned subsets of items. Our goal is to design incentives for truthful ordinal reporting without requiring agents to reveal their full private information. To do so, we leverage the fact that ordinal reports naturally encode pairwise comparisons among items.

This suggests a simple agreement-based mechanism: reward agents according to whether their reported pairwise comparisons agree with those of their peers, equivalently through a Kendall tau-based pairwise-agreement score. Such an approach is attractive because it is simple, transparent, and directly aligned with the ordinal information the designer wants to elicit. In general elicitation settings, however, rewarding agents for agreement can encourage them to suppress private information when they believe their signal is likely to be in the minority~\cite{kong2016putting}. For example, if an evaluator believes that most peers will mark nearly every submission as acceptable, then a reward based only on agreement may induce the evaluator to report ``acceptable'' even for a submission they privately judge to be weak. This is why agreement mechanisms typically require strong self-dominance conditions---after observing a signal, an agent must believe that other agents are more likely to report the same signal than any alternative~\cite{zhang2025stochastically}---to have a truthful reporting equilibrium.

Our key observation is that, although such self-dominance conditions are generally strong for arbitrary signals, they become natural at the level of pairwise ordinal comparisons. We identify a broad class of Bayesian Stochastically Order Consistent (Bayesian SOC) information structures under which each truthful pairwise comparison is more likely to agree with a peer's comparison than to disagree, even after conditioning on the agent's richer private information. Under this condition, the Kendall tau mechanism makes truthful ordinal reporting stochastically dominant, despite eliciting only a coarse ordinal summary of the agent's information.

\paragraph{Contributions.}
We introduce a peer-prediction mechanism for coarse ordinal reports based on Kendall tau distance and its pairwise-agreement extension. The mechanism applies to several natural reporting formats, including full rankings, fixed-shape bucket orders~\cite{fagin2004comparing}, and rankings over assigned subsets of items. The bucket-order setting covers full rankings and top-$K$ sets as special cases and also captures tiered reports. Its reward rule is simple and transparent: agents are paid according to agreement with peers on the pairwise comparisons induced by their reports. This simplicity is useful both theoretically and practically. When evaluators are already comparing a small set of alternatives, ordinal reports can be easier to specify and audit than calibrated cardinal scores, consistent with a long line of work in psychometrics on comparative judgment~\cite{thurstone2017law,tarricone2016using,marshall2020assessment}.

We provide strong incentive guarantees for this mechanism under a class of Bayesian Stochastically Order Consistent (Bayesian SOC) information structures. Bayesian SOC ensures that each truthful pairwise comparison is more likely to agree with a peer's comparison than to disagree, even after conditioning on the agent's richer private information. This class includes classical ranking information structures, such as the Thurstone and Mallows models. Under Bayesian SOC, truthful ordinal reporting is stochastically dominant: truth-telling not only maximizes expected payment, but yields a payment distribution that first-order stochastically dominates the distribution under any unilateral deviation. This is stronger than standard equilibrium guarantees based only on expected utility and provides robustness to agents with nonlinear utility functions, such as risk aversion.  Moreover, we show that the mechanism is strongly truthful if agents' strategies cannot coordinate on an external indexing of items.

We also study the measurement accuracy of the mechanism, namely how well payments track the quality of an agent's report \cite{chen2025data}. This property is important in practice for fairness~\cite{burrell2022measurement}, motivation~\cite{ZhangS2023Higheffort}, and trust in the mechanism. Prior work has shown that strong stochastic-dominance incentives can come at a substantial cost to accuracy~\cite{zhang2025stochastically}. In contrast, our experiments on synthetic data from Mallows and Thurstone models, together with a real-data stress test on the SUSHI preference dataset~\cite{kamishima_sushi_dataset}, suggest that the Kendall tau mechanism remains sensitive to report quality in the ordinal settings we study.

\subsection{Relation to Prior Work}\label{sec:related} Peer prediction began with mechanisms such as Bayesian Truth Serum~\cite{prelec2004bayesian} and has developed into a broad literature on eliciting unverifiable information from correlated reports. A central difficulty is that simple agreement can make agents coordinate on common or focal reports rather than reveal their private information~\cite{kong2016putting}. Many later mechanisms therefore reward reports through prediction, correlation, or mutual-information-style quantities, including informed-truthful and multi-task mechanisms~\cite{shnayder2016informed,kong2019information,kong2020dominantly,schoenebeck2020learning,feng2022peer}. These mechanisms typically operate on categorical or numerical reports, often require agents to report a sufficiently rich representation of their signal, or learn/estimate statistical relationships across tasks (items). Our mechanism instead asks for ordinal reports and uses a fixed, transparent pairwise-agreement score.

Our setting connects most directly to the bonus-penalty payment (BPP) mechanism for binary pairwise comparisons~\cite{chen2024carrot}, which operates under Bayesian Strong Stochastic Transitivity (Bayesian SST). BPP has a three-agent structure: an agent evaluates one pair and is scored using signals from two other agents on related pairs. This makes BPP well suited to sparse pairwise-comparison assignments. By contrast, our mechanism asks agents for richer ordinal reports---full rankings, top-$K$ sets, bucket orders, and subset assignments---and scores them by all-pairs agreement on overlapping item pairs. Thus our framework is less general in assignment structure, because pairs must be reviewed by more than one agent, but it handles richer reports directly.

The modeling assumptions are also different. BPP is designed for homogeneous pairwise signals, whereas Bayesian SOC allows heterogeneous agents and full signal vectors.  For instance, different reviewers may have different noise levels in a Thurstone or Mallows model. The two frameworks are not compatible. \Cref{app:discussion} compares SOC with SST in detail.

Our stochastic-dominance guarantee is related to recent stochastically dominant peer prediction for multi-task categorical reports~\cite{zhang2025stochastically}. Their mechanisms use task partitions to obtain stochastic dominance for categorical signals. Here, the ordered structure of rankings and Bayesian SOC allow a simpler Kendall tau-based pairwise-agreement score: the proof turns a deviation into adjacent pair corrections, each improving the score distribution. Technically, the proof controls weighted peer-alignment margins rather than independent categorical tasks.

Conceptually, our framework also connects to property elicitation~\cite{lambert2008eliciting}: the target function maps a rich private signal vector to the ordinal property we want to elicit, such as a ranking, top set, or bucket order. From this perspective, Kendall tau-based pairwise agreement is a scoring rule for an ordinal projection of the agent's information rather than for the full signal. This distinction is important in settings such as review, peer assessment, and RLHF, where the platform often wants a tractable ordinal judgment even though agents may observe richer cardinal evidence.

\paragraph{Paper organization.} \Cref{sec:pre_rank} introduces the Bayesian SOC framework and \cref{sec:ppm} formalizes ordinal peer prediction. \Cref{sec:full} proves the full-ranking Kendall tau mechanism is truthful and stochastically dominant truthful. \Cref{sec:extensions} gives the subset-assignment and bucket-order extensions, the latter also implying the top-$K$ result. \Cref{sec:exp} reports the main empirical comparison; proofs, specialized top-set details, and additional experiments are in the appendices.

\section{Problem Formulation} \label{sec:problem}
We consider a collection of $M$ items and a set of $N\ge2$ strategic agents. Each agent privately observes noisy signals for the items. Our goal is to design mechanisms to elicit agents' ordinal information from private signals truthfully. We introduce the joint distributions of agents' private signals in~\Cref{sec:pre_rank} and then define the information elicitation problem in~\Cref{sec:ppm}.

\nomenclature{$N$}{The number of agents}
\nomenclature{$M$}{The number of items (candidates)}
\nomenclature{$n,n'$}{indices of agent and her peer}

\nomenclature{$i,j$}{indices of items}
\nomenclature{$k,l$}{ranks of items}

We first define Kendall tau distance between two item rankings. Given two items $i$ and $j\in [M]$ and an ordering $r:[M]\to [M]$ of them, the sign function is
$$\sign(r(i),r(j)) = \begin{cases} 1&\text{ if }r(i)<r(j)\\
    -1&\text{ if }r(i)>r(j)\\
    0&\text{ otherwise}
    \end{cases},$$
where $r(i)$ is the rank of the item $i$ in the ordering $r$. For notation brevity, given two rankings, $r$ and $r'$, we further set $$\sign(r(i),r(j); r'(i),r'(j)) = \sign(r(i),r(j))\sign(r'(i),r'(j)).$$ We say $r$ and $r'$ are order consistent at $(i,j)$ if $\sign(r(i), r(j); r'(i), r'(j))\ge 0$, and strictly order consistent if $\sign(r(i), r(j); r'(i), r'(j))>0$.

\begin{definition}\label{def:kendall_tau}
    Given two full rankings $r, r' \in \symm_M$, the \emph{Kendall tau distance} between them counts how many pairs of items are ordered differently in $r$ and $r'$. That is,
    \begin{equation}\label{eq:kendall1}
       d_\tau(r, r') := \frac{1}{2} \sum_{i<j} \left(1 - \sign\left(r(i), r(j); r'(i), r'(j)\right)\right).
    \end{equation}
\end{definition}

\subsection{Bayesian Stochastically Order Consistent Models}\label{sec:pre_rank}

Given the set of $M$ items, each item $i\in [M]$ has an underlying unknown state $\theta_i \in \Theta$, where $\Theta$ is a (strictly) ordered set defined in~\cref{app:pre}. The vector of states $\vtheta = (\theta_1,\dots,\theta_M)$ is drawn from a prior $\mu$. Depending on the model, $\vtheta$ may represent either a reference ranking or underlying item costs.

For each item, each agent gets a signal from a (strictly) ordered set $\mathcal{X}$.  Conditioned on a realized $\vtheta$, each agent $n\in [N]$ independently receives a private signal vector $\vx_n\in \mathcal{X}^M$ from a distribution $P_n(\cdot\mid \vtheta)$ over these vectors of signals.
Let $\rmX = (\rvx_n)_{n\in [N]}$ be the random matrix of \emph{all} agents' signals, where each agent $n$ receives a signal vector $\rvx_n = (\rx_{n,1},\dots,\rx_{n,M})$. When realized, these values are denoted as $\mX$ for the matrix, $\vx_n$ for the row of agent $n$, and $x_{n,i}$ for the entry for agent $n$ and item $i$.

An \emph{information structure} consists of a prior $\mu$ on $\vtheta$ and the collection of conditional signal distributions $\mP = (P_n)_{n\in [N]}$.
Given $i,j\in [M]$ and strictly ordered sets $\mathcal{Y}, \mathcal{Z}$,  $\vy\in \mathcal{Y}^M, \vz\in \mathcal{Z}^M$ are \emph{order consistent at $(i,j)$} if their relative orderings at those indices match.

\begin{definition}[Kleinberg et al.~\cite{kleinbergprice}]\label{def:soc}
    Given $M$, and ordered sets $\Theta$ and $\mathcal{X}$, a conditional probability $P:\Theta^M\to \Delta(\mathcal{X}^M)$ is \emph{stochastically order consistent} (SOC) if for any distinct $i,j\in [M]$, state vector $\vtheta\in \Theta^M$, and signal vector $\vx\in \mathcal{X}^M$ which 
    are order consistent at $(i,j)$,
    $$P(\rvx = \vx\mid \vtheta)\ge  P(\rvx = \sigma_{ij}(\vx)\mid  \vtheta)$$ where $\sigma_{ij}(\vx)$ denotes the vector $\vx$ with the $i$-th and $j$-th components swapped.\footnote{If the signal space $\mathcal{X}^M$ is continuous or more general, we assume the conditional probability is absolutely continuous with respect to a common measure invariant under coordinate permutations, and the inequality holds for the probability density function.}  Moreover, the inequality is strict when $\vtheta$ and $\vx$ are strictly order consistent at $(i,j)$.

    \nomenclature{$\sign$}{sign function}
\end{definition}
Intuitively, for any two items $i$ and $j$, it is more probable that their order in the signal is consistent with their true order in $\vtheta$ than the reverse, regardless of the realizations of the other items.  We discuss the connection to order independence in~\cref{app:discussion}.

\begin{definition}[{Bayesian stochastically order-consistent}]\label{def:bsoc}  An information structure $(\mu, \mP)$ is \defn{Bayesian stochastically order-consistent} (Bayesian SOC) if it satisfies:
\begin{itemize}
\item Ex-ante identical: $\vtheta$ is drawn from an exchangeable prior $\mu$.\nomenclature{$\mu$}{Prior on the state of each item}
    \item Neutrality: For each agent $n$, the conditional probability $P_n$ is invariant to permutations of the items.  For all permutations $\sigma\in \symm_M$, $\vx$, and $\vtheta$, $P_n(\vx\mid \vtheta) = P_n(\sigma(\vx)\mid \sigma(\vtheta))$.
    \item Stochastic consistency: Each conditional signal distribution $P_n$ is SOC (\cref{def:soc}), so the signals are more likely to be order consistent with the true underlying order.
\end{itemize}
A Bayesian SOC model is \emph{strict} if there are no ties: $\Pr[\rx_{n,i} = \rx_{n,j}] = 0$ and $\Pr[\theta_i = \theta_j] = 0$ for all $n\in [N]$ and $i\neq j\in [M]$. Finally, $\mP$ has \emph{full support} if $P_n(\cdot\mid \vtheta)$ assigns positive probability to the same set of signals for all $n$ and $\vtheta$.
\end{definition}

As shown in~\cref{ex:mallows,ex:para,ex:rum}, several classical ranking models embody this setting and satisfy Bayesian SOC, including the Mallows model~\cite{1379e110-3390-3287-8a9b-6ccafc65156a}, the Thurstone model~\cite{thurstone2017law}, and additive random utility models~\cite{thurstone2017law,DBLP:conf/nips/SoufianiPX12,block1959random}.

\begin{example}[Mallows model~\cite{1379e110-3390-3287-8a9b-6ccafc65156a}]\label{ex:mallows} 
Here the vector of item states $\vtheta$ and agent $n$'s signal vector $\vx_n$ are each rankings of the items. The signal for each item is then its order in the ranking, so $\Theta = \mathcal{X} = [M]$. Because the full signal is a permutation, the vectors lie in the smaller set $\symm_M \subset \Theta^M = \mathcal{X}^M$. For the \emph{Mallows model} with $\veta$, $\vtheta\in \symm_M$ is a reference ranking (e.g., the ground-truth ordering of items) and $\veta = (\eta_1,\dots,\eta_N)\in \R_{>0}^N$ is a noise parameter vector, allowing each agent to have their own ``ranking noise.'' 

In this model, agent $n$'s observed ranking $\vx\in \symm_M$ is generated with probability
\[
P_n(\rvx_n = \vx\mid \vtheta) \propto \exp\left(-\frac{1}{\eta_n} d_\tau(\vtheta, \vx)\right)
\]
where $d_\tau$ is Kendall's tau distance (see~\cref{def:kendall_tau}), which measures the number of pairwise disagreements between two rankings.
\end{example}
    
\noindent In the Mallows model, smaller $\eta_n$ yields rankings closer to the reference $\vtheta$, while larger $\eta_n$ produces noisier (more random) rankings.

\begin{example}[Thurstone model~\cite{thurstone2017law}]\label{ex:para}
Let $\Theta = \mathcal{X} = \R$. The latent state $\vtheta$ is a vector in $\R^M$, each coordinate $\theta_i$ being the true cost of item $i$. We assume an exchangeable prior with pairwise distinct costs almost surely, for example, i.i.d.\ standard Gaussian costs.

Each agent $n$ then sees a noisy version of each $\theta_i$: they observe Gaussian signals $\rx_{n,i}$, where
\[
\rx_{n,i}\mid\vtheta \sim \mathcal{N}(\theta_i, \eta_n)
\]
with $\eta_n>0$ specifying the noise level for agent $n$. Conditional on $\vtheta$, the signals are independent across items and agents.

\end{example}
\noindent In the Thurstone model, each agent sees noisy measurements of the underlying item costs, with more reliable agents having lower $\eta_n$. The Thurstone model is a classical foundation for statistical models of ranking and paired comparison.  While agents ultimately report an ordering (ranking) derived from these perturbed costs, unlike the Mallows model, in the Thurstone model the agent signal is not a ranking per se. 
    
\begin{example}[Additive RUM~\cite{thurstone2017law,DBLP:conf/nips/SoufianiPX12,block1959random}]\label{ex:rum}
Let $\Theta = \mathcal{X} = \mathbb{R}$. The true item costs $\vtheta\in\mathbb{R}^M$ have an exchangeable prior and are pairwise distinct almost surely. Each agent $n$ observes
\[
\rx_{n,i} = \theta_i + \epsilon_{n,i}
\]
for each item $i$. The noise terms are independent of $\vtheta$ and of each other, and for each agent they have a common everywhere-positive, strictly log-concave density on $\mathbb{R}$. Examples include Gaussian and reflected Gumbel noise. With reflected Gumbel noise, ranking costs in increasing order is equivalent to ranking the utilities $-\theta_i$ plus Gumbel noise in decreasing order, yielding a Plackett--Luce ranking model~\cite{luce1959individual,plackett1975analysis}.
\end{example}
\noindent
For $\theta_i<\theta_j$ and $x_i<x_j$, the residuals $x_i-\theta_i$ and $x_j-\theta_j$ lie strictly between the swapped residuals $x_i-\theta_j$ and $x_j-\theta_i$ and have the same sum. Strict concavity of the log density therefore gives a strictly larger product density for the aligned observations. This establishes the SOC swap inequality. The Thurstone model is the Gaussian special case.

Note that agents can be heterogeneous when $P_n\neq P_{n'}$.  Throughout the paper, for readability, we use probability-mass notation such as $P(\rvx=\vx\mid\vtheta)$ and write sums over states or signals. In continuous state or signal spaces, these statements and calculations are interpreted as the relevant densities and integrals under common dominating measures.

Thus, our general requirement is satisfied by several standard ranking models, rather than by a mechanism-specific prior.  In~\cref{app:discussion}, we discuss its connection to the Independence of Irrelevant Alternatives principle and give a necessity result illustrating why the SOC orientation matters for agreement-based incentives.

\subsection{Peer Prediction Mechanism Design for Ordinal Data}\label{sec:ppm}

We want to elicit ordinal information from agents.  For each agent $n$, we formalize this via a \emph{target function} $\Gamma_n: \mathcal{X}^M \to \mathcal{R}_n$, which maps a signal vector to a specific (truthful) report.\footnote{To simplify the analysis, we assume the information structure is a strict Bayesian SOC model (see~\cref{sec:pre_rank}), which ensures distinct signal values and a unique target function.}

\begin{description}
    \item[Full ranking] $\mathcal{R}_n = \symm_M$ and $\Gamma_n = ord(\vx)$.  The target function $\text{ord}(\vx)$ outputs a permutation $r$ with $r(i)<r(j)$ if and only if $x_i< x_j$  (lower signals are better).
    \item[Top set] Given a size parameter $K_n\in \mathbb{Z}_{>0}$, $\mathcal{R}_n = \binom{[M]}{K_n}$ and $\Gamma_n = top_{K_n}(\vx)$. The target function outputs the indices of the $K_n$  smallest signals.
    \item[Bucket order]~\cite{fagin2004comparing} Given a shape parameter $\bm{K}^n = (K^n_1,\dots,K^n_{L_n})$ with $\sum_{\ell} K^n_\ell = M$, $\mathcal{R}_n$ is the set of ordered partitions of $[M]$ into buckets $B_1,\dots,B_{L_n}$ with $|B_\ell| = K^n_\ell$. The target function $\Gamma_n$ places the $K^n_1$ items with the smallest signals in $B_1$, the next $K^n_2$ in $B_2$, and so on.  Full rankings and top sets are the shapes $(1,\dots,1)$ and $(K_n, M-K_n)$ defined in~\cref{def:bucketorder}.
    \item[Subset assignment] Given an assigned subset $S_n \subseteq [M]$, the target function $\Gamma_n$ is the ranking $\text{ord}(\vx)$ restricted to these items. For instance, an assigned pair, $|S_n| = 2$, reproduces the setting in~\citet{chen2024carrot}; see~\cref{sec:subset}.
\end{description}

To incentivize agents to truthfully report the target function of their private signals, a peer prediction mechanism creates a game between the agents.  Formally, given {target functions} $(\Gamma_n)_{n\in [N]}$, each agent $n$ privately observes her signal $\vx_n\in \mathcal{X}^M$, and reports $\tilde{r}_n\in \mathcal{R}_n$ which is potentially different from the truthful target report $r_n := \Gamma_n(\vx_n)$.
A \emph{peer prediction mechanism} $(U_n)_{n\in [N]}$ takes all agents' reports $\tilde{\vr} = (\tilde{r}_1,\dots,\tilde{r}_N)$ and pays/scores agent $n$ with $U_n(\tilde{\vr})\in \R$.

Each agent $n$'s strategy is a random function from her signal to a report $\alpha_n: \vx_n\mapsto \tilde{r}_n$, and the randomness of their strategies is independent of each other's and all signals.   A (mixed) strategy profile $\valpha = (\alpha_1,\dots,\alpha_N)$ is a collection of all agents' strategies.  All agents are rational. 
\nomenclature{$\alpha_n$}{Agent $n$'s (mixed) strategy}

\begin{definition}
    A mechanism is \defn{truthful} if the truthful strategies form a Bayesian Nash equilibrium: for every agent $n$, signal $\vx_n$, and strategy $\alpha_n$ with $\Pr[\alpha_n(\vx_n)\neq \Gamma_n(\vx_n)]>0$,
 $$\E[U_n(\Gamma_n(\vx_n), {\rvr}_{-n})\mid \vx_n]>\E[U_n(\alpha_n(\vx_n), {\rvr}_{-n})\mid \vx_n].$$
 Here the deviation probability is over the randomness of $\alpha_n$ at the fixed signal $\vx_n$. The expectation is over the peers' truthful reports $\rvr_{-n} := (\Gamma_{n'}(\rvx_{n'}))_{n'\neq n}$ and the randomness of $\alpha_n$.
\end{definition}

Beyond expected utility, we also consider a stronger property where the truthful score distribution stochastically dominates the score distribution of any deviation.
\begin{definition}[\cite{zhang2025stochastically}]\label{def:sdt}
    A mechanism is \defn{stochastically dominant truthful} (SDT) if for every agent $n$, signal $\vx_n$, and deviating strategy $\alpha_n$ with $\Pr[\alpha_n(\vx_n)\neq \Gamma_n(\vx_n)]>0$, where this deviation probability is over the strategy randomness at the fixed signal, the truthful payment random variable first-order stochastically dominates the payment from the deviation when the peers report truthfully. Formally, for all $w \in \mathbb{R}$:
\begin{equation*}
    \Pr[U_n(\Gamma_n(\vx_n), {\rvr}_{-n}) \ge w \mid \vx_n] \ge \Pr[U_n(\alpha_n(\vx_n), {\rvr}_{-n}) \ge w \mid \vx_n]
\end{equation*}
with strict inequality for at least one $w\in\mathbb{R}$. The randomness is over the peers' truthful reports $\rvr_{-n}$, the randomness of $\alpha_n$, and any internal randomness of the mechanism.
\end{definition}

Traditional truthful mechanisms require agents to be risk neutral and for their \emph{utilities} to be linear with respect to their \emph{scores}.  With these assumptions, if the expected score decreases, so does their utility.  However, these assumptions are often unrealistic.  For example, students 
typically care about letter grades rather than numerical scores; in a crowd work setting, there may be a minimum score needed for rehiring.  SDT addresses this by guaranteeing that the score distribution from truth-telling first-order stochastically dominates any deviation, incentivizing truthfulness for all monotone utility functions. This robustness makes SDT mechanisms suitable for practical applications where agents' utilities may be unknown or nonlinear and where agents may not be risk neutral, such as peer assessment or human feedback for AI.   

Truthfulness and SDT are unilateral guarantees: they compare truth-telling with a single agent's deviation while peers report truthfully. They do not rule out other profiles, such as all agents reporting the same fixed ranking, that pay as much as truth-telling. Strong truthfulness, adapted from~\cite{shnayder2016informed}, instead compares truth-telling with every strategy profile in a class $\mathcal{A}$. For full rankings, let $rev\circ r(i) := M+1-r(i)$ denote the global reversal of a ranking.
\begin{definition}\label{def:strong}
    A mechanism is \defn{strongly truthful} over a class $\mathcal{A}$ of strategy profiles if for every $\valpha\in\mathcal{A}$ and every agent $n$,
    \begin{equation*}
        \E\bigl[U_n(\Gamma_1(\rvx_1),\dots,\Gamma_N(\rvx_N))\bigr] \ge \E\bigl[U_n(\alpha_1(\rvx_1),\dots,\alpha_N(\rvx_N))\bigr],
    \end{equation*}
    where the expectation is ex ante over all signals and strategy randomness, with equality only if $\valpha$ is the truthful profile or its global reversal $(rev\circ\Gamma_n)_{n\in[N]}$.
\end{definition}
\Cref{sec:strong} takes $\mathcal{A}$ to be the index-equivariant profiles.

\section{Kendall Tau Mechanisms for Eliciting Full Rankings}\label{sec:full}

We introduce the Kendall tau mechanism for full rankings in \cref{alg:kendall_tau}, which penalizes agents based on their Kendall tau distance from peers.  We add the constant $\binom{M}{2}$ to simplify the notation. Since Kendall tau distance treats a ranking as a collection of all pairwise comparisons, the mechanism rewards agents for every pair $(i,j)$ whose order is consistent with a peer's order. Thus, we may view Kendall tau mechanisms as agreement mechanisms on pairwise comparisons.

\begin{algorithm}[t]
\caption{Kendall Tau Mechanism for Full-Rank Data}
\label{alg:kendall_tau}

\KwIn{$\tilde{\vr} \in \symm_M^N$, the reported full rankings of $N$ agents over $M$ items}

\ForEach{agent $n\in [N]$}{
    Compute score
    \begin{equation}\label{eq:score_full}
        U_n(\tilde{\vr})
        =
        \sum_{n' \neq n} \left(\binom{M}{2}- 2d_\tau(\tilde{r}_n, \tilde{r}_{n'})\right) = \sum_{n'\neq n}\sum_{i<j} \sign(\tilde{r}_n(i),\tilde{r}_n(j); \tilde{r}_{n'}(i), \tilde{r}_{n'}(j)).
    \end{equation}
}
\end{algorithm}

As a warm-up, \cref{sec:full_truthful} proves that the Kendall tau mechanism is truthful. In particular, \cref{lem:consist} uses symmetrization to compare the probability of order consistency and inconsistency. \Cref{sec:sec:full_sd} strengthens the result to stochastically dominant truthfulness.  In particular, \cref{thm:sd} implies \cref{thm:truthful}. Finally, \cref{sec:strong} shows the mechanism is strongly truthful under index-equivariant strategies.  All deferred proofs are in~\cref{app:full,app:strong}.

\subsection{Truthfulness}\label{sec:full_truthful}
 \begin{proposition}\label{thm:truthful}
     For any strict Bayesian SOC information structure, the Kendall tau mechanism in~\cref{alg:kendall_tau} is truthful.
 \end{proposition}

 The following lemma shows that under a Bayesian SOC model, the conditional probability of any peer's signals given the agent's private signal is also stochastically order consistent. 
 \begin{lemma}\label{lem:consist}
     If $(\rvx,\rvx')$ are sampled from a strict Bayesian SOC model, then for any distinct $i,j\in[M]$ and signals $\vx,\vx'\in\mathcal{X}^M$ that are strictly order consistent at $(i,j)$, with $\vx$ of positive marginal mass or density,
     $$\Pr[\rvx' = \vx'\mid \rvx = \vx]>\Pr[\rvx' = \sigma_{i,j}(\vx')\mid \rvx = \vx]$$
     where $\sigma_{i,j}$ is the permutation that swaps coordinates $i$ and $j$.
 \end{lemma}

\Cref{thm:truthful} follows immediately from the pairwise posterior-agreement property in~\cref{lem:consist}: the expected utility decomposes pairwise, and each pair is maximized by the truthful comparison. We defer the proof to~\cref{app:full}.

To analyze the full mechanism and facilitate the extensions, we define a comparison function from full rankings and rewrite Kendall tau distance.
 
 \begin{definition}[Comparison function]\label{def:comp} Let $comp:\symm_M \to \{-1,0,1\}^{M\times M}$ output comparison matrix $\mD = comp(r)$ for any ranking $r\in \symm_M$ so that for all $i,j$
$$\mD(i,j) = \sign(r(i),r(j)).$$
We can further define the comparison function on partial orders $R\subset[M]^2$ so that $\mD(i,j) = \vone[(i,j)\in R]-\vone[(j,i)\in R]$.
 \end{definition}
With this notation, the Kendall tau distance between $r$ and $r'$ is  $d_\tau(r,r') =
\frac{1}{2}\sum_{i<j}(1-\mD(i,j)\mD'(i,j)),$
where $\mD = comp(r)$ and $\mD' = comp(r')$, and the score in~\cref{eq:score_full} becomes
\begin{equation}\label{eq:kendall_comp}
    U_n(\tilde{\vr}) = \sum_{n'\neq n}\sum_{i<j}\tilde{\mD}_n(i,j)\tilde{\mD}_{n'}(i,j)
\end{equation}
where $\tilde{\mD}_n:= comp(\tilde{r}_n)$ for all $n$. Similarly, define $\mD_n := comp(r_n)$ for the truthful report of agent $n$. 
\nomenclature{$\mD$}{Comparison matrix}

\subsection{Stochastically Dominant Truthfulness}\label{sec:sec:full_sd}

\begin{theorem}\label{thm:sd}
    For any strict Bayesian SOC information structure, the Kendall tau mechanism 
    in~\cref{alg:kendall_tau} is stochastically dominant truthful.
\end{theorem}

\paragraph{Proof overview.}
To prove~\cref{thm:sd}, we show that if all other agents are truthful, deviating to any nontruthful report $\tilde{r}$ yields a first-order stochastically dominated score. Specifically, given agent $n$ and her private signal $\vx_n$, we construct a sequence of reports $(r^t)_{t=0,\dots,T}$ for agent $n$ with payments $U^t = U_n(r^t, \Gamma(\mX_{-n}))$, where $\Gamma(\mX_{-n}) := (\Gamma_{n'}(\vx_{n'}))_{n'\neq n}$ denotes the peers' truthful reports, so that $r^0=\tilde{r}$, $r^T=\Gamma_n(\vx_n)$ is the truthful report, and $$U^t\ge_{\mathrm{FOSD}}U^{t-1}\quad\text{for all }t=1,\dots,T,$$
which completes the proof by transitivity of first-order stochastic dominance. We construct the sequence by bubble sort.

\begin{definition}[Correction path]\label{def:path}
    Given an agent's signal $\vx$ and a reported ranking $\tilde{r}$, we define the \emph{correction path} as the sequence of rankings $r^0, r^1, \dots, r^T$ to transform $\tilde{r}$ into $ord(\vx)$ through the bubble-sort algorithm.
    \begin{enumerate}
        \item Start: $r^0 = \tilde{r}$.
        \item Step $t$: Given $r^{t-1}$, we find a rank $k_t$ such that the item at rank $k_t+1$ (called $i_t$) and the item at rank $k_t$ (called $j_t$) satisfy $x_{i_t} < x_{j_t}$---out of order relative to the signal. We then swap their ranks, setting $r^t(i_t) = k_t$ and $r^t(j_t) = k_t+1$.
        \item End: $r^T = ord(x)$, where no such adjacent contradictions exist.
    \end{enumerate}
    Finally, let $U^t(\mX_{-n}) = U_n(r^t, \Gamma(\mX_{-n}))$ denote agent $n$'s payment at step $t$ given peer signals. Each step removes exactly one inversion of $r^{t-1}$ relative to $ord(\vx)$, so $T$ equals the number of inversions of $\tilde{r}$, and $T\le\binom{M}{2}$.
\end{definition}

Because $i_t$ and $j_t$ are adjacent in both $r^{t-1}$ and $r^t$, the only
comparison that changes is the one between $i_t$ and $j_t$. We therefore can decompose the peers' signals into \emph{magnitude} $\rho$, which specifies everything except how each peer orders this pair, and  \emph{alignment} $\va=(a_{n'})_{n'\neq n}\in\{-1,1\}^{N-1}$, where $a_{n'}=1$ if peer $n'$ orders $(i_t,j_t)$ consistent with $r^t$.

\begin{definition}[Magnitude, alignment vectors, and margins]\label{def:events}
Given an item pair $(i_t, j_t)$ and a reference agent $n$ with signal vector $\vx_n$ so that $x_{n,i_t}<x_{n,j_t}$, we classify the outcome space of peers' signal matrix $\mX_{-n} \in \mathcal{X}^{(N-1)\times M}$ into equivalence classes $\mathcal{V}$ called \emph{magnitude}.
We say two matrices $\mX_{-n}$ and $\mX'_{-n}$ have the same magnitude if for every peer $n' \neq n$: \nomenclature{$\rho$}{Magnitude of peers' signal matrices}

\begin{enumerate}
    \item The set of entries at the pair indices is identical: $\{x_{n', i_t}, x_{n', j_t}\} = \{x'_{n', i_t}, x'_{n', j_t}\}$;
    \item All other entries are identical: $x_{n', k} = x'_{n', k}$ for all $k \notin \{i_t, j_t\}$.
\end{enumerate}
For any $\rho\in \mathcal{V}$, let $\hat{\mX}_{-n}$ be the canonical element so that each peer's signal $\hat{\vx}_{n'}$ and $\vx_n$ are order consistent at $(i_t,j_t)$, with $\hat{x}_{n', i_t} < \hat{x}_{n', j_t}$.
For any matrix $\mX_{-n} \in \rho$, we define the \emph{alignment vector} $\va=(a_{n'})_{n'\neq n}\in \{-1, 1\}^{N-1}$ to indicate whether each peer is order consistent with agent $n$. For all $n'\neq n$,
$$a_{n'} = \sign(x_{n', i_t}, x_{n', j_t};x_{n, i_t},x_{n, j_t}).$$
Given $\rho$ and $\va$ with canonical element $\hat{\mX}_{-n}$,
\begin{equation}
    x_{n',i} = \begin{cases}
        \hat{x}_{n',j_t} & \text{if } i = i_t \text{ and } a_{n'} = -1 \\
        \hat{x}_{n',i_t} & \text{if } i = j_t \text{ and } a_{n'} = -1 \\
        \hat{x}_{n',i} & \text{otherwise}.
    \end{cases}
\end{equation}
We denote the magnitude and alignment vector of a matrix by $\rho(\mX_{-n})$ and $\va(\mX_{-n})$. We define $\check{\mX}_{-n}$ as the fully reversed configuration corresponding to $\va = -\mathbf{1}$.\nomenclature{$\va$}{Alignment vector}

Finally, the \emph{margin} is $L(\va) := \sum_{n'\neq n} a_{n'}$.\nomenclature{$L_B$}{The margin function on a subset of peers $B$}
\end{definition}

Let $Q(\rho,\va)$ be the joint conditional mass of $(\uprho,\rva)$ given $\vx_n$ at $(\rho,\va)$. In the density setting, $Q$ is a density in $\rho$ and a mass in $\va$, using the product reference measure restricted to canonical representatives. Sums over $\rho$ then denote integrals. Let $U^t(\rho,\va)=U_n(r^t,\Gamma(\mX_{-n}))$ be agent $n$'s payment at step $t$ given peer signals parameterized by $(\rho,\va)$.

\Cref{lem:pair1} shows that, at a fixed magnitude, the score difference depends on the alignment only through the \emph{margin} $L(\va)$. Write $Q(\rho,L=d)=\sum_{\va:L(\va)=d}Q(\rho,\va)$.

\begin{lemma}\label{lem:pair1}
For each adjacent correction step, fixing the magnitude $\rho$, the score difference is determined by the peer-alignment margin. In particular, for any alignment vectors $\va,\va'$,
\begin{equation}\label{eq:sd_pair11}
    U^t(\rho, \va)-U^t(\rho, \va') = L(\va)-L(\va')
\end{equation}
and
\begin{equation}\label{eq:sd_pair12}
    U^t(\rho,\va) = U^{t-1}(\rho,\va) + 2L(\va).
\end{equation}
\end{lemma}

\begin{lemma}\label{lem:pair2}
Under strict Bayesian SOC, for any agent signal $\vx_n$ of positive marginal mass or density, magnitude $\rho$, and $d>0$,
\begin{equation}\label{eq:sd_pair3}
    Q(\rho,L=d)\ge Q(\rho,L=-d).
\end{equation}
\end{lemma}

By~\cref{lem:pair1}, each correcting swap changes the score by a monotone function of the margin. Pairing opposite margins using~\cref{lem:pair2} gives one-step weak stochastic dominance. We compose these steps and prove strictness separately via expected payments. The proofs of the two lemmas are in~\cref{app:full}.

\begin{proof}[Proof of \cref{thm:sd}]
Given $\rvx_n = \vx_n$ and a correction path $r^0,\dots,r^T$ as in~\cref{def:path}, it is enough to show
\begin{equation}\label{eq:sd1}
    U^t\ge_{\mathrm{FOSD}}U^{t-1}, \qquad \forall t = 1,\dots,T.
\end{equation}
Then transitivity of stochastic dominance gives, for all $w\in \R$,
\begin{align*}
    &\Pr[U_n(ord(\vx_n), \Gamma(\rmX_{-n}))\ge w\mid \vx_n]= \Pr[U^{T}\ge w\mid \vx_n]\\
    \ge& \Pr[U^{0}\ge w\mid \vx_n]\tag{by \cref{eq:sd1}}\\
    =& \Pr[U_n(\tilde{r}, \Gamma(\rmX_{-n}))\ge w\mid \vx_n].
\end{align*}
To prove~\cref{eq:sd1}, by~\cref{eq:sd_pair11}, define $u(\rho,d) = U^t(\rho,\va)$ for all $\rho$ and $\va$ with $L(\va) = d$. Then
\begin{align*}
    &\Pr[U^t(\uprho, \rva)\ge w\mid \vx_n]-\Pr[U^{t-1}(\uprho, \rva)\ge w\mid \vx_n]\\
    =& \sum_{\rho, d}\sum_{L(\va) = d}Q(\rho,\va)\left(\vone[U^t(\rho,\va)\ge w]-\vone[U^{t-1}(\rho,\va)\ge w]\right)\\
    =& \sum_{\rho, d}\sum_{L(\va) = d}Q(\rho,\va)\left(\vone[u(\rho,d)\ge w]-\vone[u(\rho,d)-2d\ge w]\right)\tag{by \cref{eq:sd_pair12}}\\
    =& \sum_{\rho, d}Q(\rho,L = d)\left(\vone[u(\rho,d)\ge w]-\vone[u(\rho,d)-2d\ge w]\right).
\end{align*}
The terms with $d=0$ cancel. Pairing positive and negative margins, the difference becomes
\begin{align*}
    &\sum_{\rho\in \mathcal{V}, d> 0}\Bigl[Q(\rho,L = d)\left(\vone[u(\rho,d)\ge w]-\vone[u(\rho,d)-2d\ge w]\right)\\
    &\qquad\qquad+Q(\rho,L = -d)\left(\vone[u(\rho,-d)\ge w]-\vone[u(\rho,-d)+2d\ge w]\right)\Bigr]\\
    =& \sum_{\rho\in \mathcal{V}, d> 0}(\vone[u(\rho,d)\ge w]-\vone[u(\rho,d)-2d\ge w])(Q(\rho,L = d)-Q(\rho,L = -d))\\
    \ge& 0,
\end{align*}
where the equality uses~\cref{eq:sd_pair11} and the final inequality uses~\cref{lem:pair2} together with $\vone[u(\rho,d)\ge w]\ge \vone[u(\rho,d)-2d\ge w]$. This completes the proof of~\cref{eq:sd1}.

Payments are bounded and weak dominance holds.  It remains to show that misreporting strictly lowers the expected payment. By the pairwise decomposition in the proof of~\cref{thm:truthful}, $\E[U_n(\tilde{r},\Gamma(\rmX_{-n}))\mid\vx_n]=\sum_{n'\neq n}\sum_{i<j}\tilde{\mD}_n(i,j)\,\E[\mD_{n'}(i,j)\mid\vx_n]$, and the strict inequality in~\cref{lem:consist} makes each $\E[\mD_{n'}(i,j)\mid\vx_n]$ nonzero with the sign of $\mD_n(i,j)$. Any ranking $\tilde{r}\neq ord(\vx_n)$ flips at least one pair, so its expected payment is strictly below the truthful one, and so is that of any strategy that misreports with positive probability.
\end{proof}

\subsection{Strong Truthfulness under Index-Equivariant Strategies}\label{sec:strong}

\Cref{thm:sd} is a unilateral guarantee and does not rule out coordinated uninformative profiles, such as every agent reporting the same fixed ranking (\cref{sec:ppm}). Such coordination, however, requires a common indexing of the items, which the designer can destroy by presenting the items to each agent under random relabeling (in RLHF, by randomly shuffling and rephrasing the responses shown to each annotator). Under randomized indexing, it is without loss of generality to restrict attention to \emph{index-equivariant} strategies.  A similar notion appears in~\cite{Zheng_2021,zhang2023multitask}.

\begin{definition}\label{def:equivariant}
A (mixed) strategy $\alpha_n : \mathcal{X}^M \to \Delta(\symm_M)$ is
\defn{index-equivariant} if
$\alpha_n(\sigma(\vx_n)) \overset{d}{=} \sigma(\alpha_n(\vx_n))$ for all
permutations $\sigma \in \symm_M$ and $\vx_n \in \mathcal{X}^M$, where
$\sigma$ acts on signal vectors by $\sigma(\vx)_i := x_{\sigma^{-1}(i)}$
and on rankings by $\sigma(r)(i) := r(\sigma^{-1}(i))$.
\end{definition}

Truth-telling $\Gamma_n = ord$ is index-equivariant, as is its composition
with the rank reversal $rev\circ r(i) := M+1-r(i)$, but a constant strategy is not.

\begin{theorem}[Strong truthfulness]\label{thm:strong}
For any strict, full-support Bayesian SOC information structure, the Kendall tau mechanism in~\cref{alg:kendall_tau} is strongly truthful (\cref{def:strong}) over index-equivariant strategy profiles.
\end{theorem}

We defer the proof to \cref{app:strong} and sketch it here. For a fixed pair $(i,j)$, we define the alignment factor $\kappa_n(\vx) := 1 - 2\Pr[\tilde{\mD}_n(i,j) \neq \sign(x_i, x_j) \mid
\rvx_n = \vx] \in [-1,1]$ where $\kappa_n = 1$ for truth-telling.  With \cref{lem:consist}, we show that the expected score can be decomposed as positive terms scaled by the product $\kappa_n \kappa_{n'} \in [-1,1]$, so truth-telling attains the maximum. 

For the equality case, full support forces each $\kappa_n$ to be a constant sign $\epsilon^{ij} \in \{\pm 1\}$ identical for all agents, and equivariance ensures the identical flip across all pairs.  Hence every agent either reports truthfully on all pairs ($\epsilon = 1$) or flips all pairs
($\epsilon = -1$), the global reversal.

\section{Extensions}\label{sec:extensions}

\subsection{Subset Assignments}\label{sec:subset}
In many practical applications, the number of items $M$ can be too large for every agent to rank. We therefore allow each agent $n$ to report only on a possibly random assigned subset $S_n\subseteq[M]$ and score only common item pairs shared with a randomly selected peer. For the comparison matrix $\tilde{\mD}_n$ induced by agent $n$'s report and one peer $n'\neq n$ drawn uniformly at random, the generalized score is
\begin{equation}\label{eq:score_subset}
    U_n(\tilde{\vr}) =
    \sum_{i,j\in S_n\cap S_{n'}:\,i<j}
    \tilde{\mD}_n(i,j)\tilde{\mD}_{n'}(i,j).
\end{equation}
For subset top-sets, where $\tilde{\mD}_n(i,j)$ records whether one item is inside the reported top set within $S_n$ and the other is outside, the peer $n'$ is instead drawn uniformly from the agents whose subsets contain $S_n$.
Before reporting, an agent observes her signal and her own assigned subset $S_n$, but not the other realized assignments or the identity of her scoring peer. The assignments are independent of states and signals. The peer is sampled after reports, using only the assignments and randomness independent of signals and reports.

Conditional on the assignment and the drawn peer, \cref{eq:score_subset} is the Kendall tau score on the shared set $S_n\cap S_{n'}$, all of which the peer observes. As first-order stochastic dominance is preserved under mixtures, the SDT guarantee is maintained after averaging over assignments and peers. To ensure that every reported comparison can affect the score, we require \emph{pairwise coverage} assignment: for every agent $n$, every realization of $S_n$, and every pair $\{i,j\}\subseteq S_n$,  $\Pr\left[\{i,j\}\subseteq S_{n'} \mid S_n\right] > 0$,
where the probability is over the assignment and the draw of $n'$.

\begin{theorem}\label{thm:subset}
For any strict Bayesian SOC information structure and any assignment satisfying pairwise coverage, the generalized Kendall tau mechanism in~\cref{eq:score_subset} is SDT for subset rankings.
\end{theorem}

Here we illustrate why~\cref{eq:score_subset} scores agent $n$ against a single random peer rather than summing over all peers. Let $M=3$ with a uniform prior over rankings, and let every agent follow a Mallows model (\cref{ex:mallows}). Agent $n$ is assigned all three items, Alice is assigned $\{1,2\}$, Bob $\{2,3\}$, and Charlie $\{1,3\}$, so every pair is covered. Alice and Charlie are accurate, and Bob's signal is close to uniformly random. Suppose agent $n$ observes $1\succ2\succ3$ and considers reporting $2\succ1\succ3$. When Alice and Charlie report the true order of their pairs, truth-telling disagrees with both of them in two true rankings, $2\succ3\succ1$ and $3\succ2\succ1$, while the misreport disagrees with both only in $3\succ1\succ2$. Therefore, truth-telling is more likely to receive the lowest payment, and does not first-order stochastically dominate the misreport.

\subsection{General Bucket Orders}\label{sec:bucket}
Bucket orders generalize full rankings, top-$K$ reports, and intermediate ordinal reports such as accept/borderline/reject tiers or ranked shortlists with unordered remainders.

\begin{definition}[Bucket order~\cite{fagin2004comparing}]\label{def:bucketorder}
A \emph{bucket order} of shape $\bm{K} = (K_1,\dots,K_L)$ with $\sum_{\ell=1}^L K_\ell = M$, $L\ge 2$, and $K_\ell>0$ for all $\ell$ is an ordered partition of $[M]$ into blocks (buckets) $B_1,\dots,B_L$ with $|B_\ell| = K_\ell$. We encode a report $r$ as $\beta_r(i)\in[L]$ for the bucket of item $i$ under $r$, with $\beta_r(i)=1$ denoting the best bucket, and its comparison matrix $\mD_r=comp(r)$ as
$$\mD_r(i,j) = \sign(\beta_r(i),\beta_r(j))\in\{-1,0,1\},$$
so $\mD_r(i,j) = +1$ if $i$ is in a strictly better bucket than $j$, and $\mD_r(i,j)=0$ if $i,j$ share a bucket.
\end{definition}

In particular, full rankings have shape $(1,\dots,1)$, and top-$K$ reports have bucket shape $(K,M-K)$. For odd $M$, a median-pivot report, which identifies the median item and separates the items above it from those below, has shape $(\frac{M-1}{2}, 1, \frac{M-1}{2})$.

We can generalize our Kendall tau mechanism to bucket orders.  The target function $\Gamma_n(\vx_n)$ sorts the signal and assigns the $K_1$ items with the smallest (best) signals to $B_1$, the next $K_2$ to $B_2$, and so on. We allow agents to have different shapes $\bm{K}^n = (K^n_1,\dots,K^n_{L_n})$ with $B^n_\ell$ for agent $n$'s buckets.  For comparison matrices, write $\langle \mA,\mB\rangle := \sum_{i<j}\mA(i,j)\mB(i,j)$. The Kendall tau mechanism pays agent $n$
\begin{equation}\label{eq:score_bucket}
U_n(\tilde{\vr}) = \sum_{n'\neq n}\langle \mD_{\tilde r_n},\mD_{\tilde r_{n'}}\rangle .
\end{equation}
\begin{theorem}\label{thm:bucket}
For any strict Bayesian SOC information structure and any per-agent shapes $\bm{K}^n$, the Kendall tau mechanism in~\cref{eq:score_bucket} restricted to shape-$\bm{K}^n$ reports is stochastically dominant truthful.
\end{theorem}

The proof mirrors the full-ranking argument but lets a correcting step move two items across adjacent bucket boundaries rather than adjacent ranks. Only comparisons involving the swapped items and the two affected buckets can change. For each peer, the score increment has the form $c_{t,n'}a_{n'}$, where $a_{n'}\in\{-1,1\}$ records alignment with the corrected pair and $c_{t,n'}\ge0$ is determined by the peer's bucket placement. Thus the one-step score improvement is a nonnegative weighted margin $\langle \vc_t,\va\rangle$. Bayesian SOC makes aligned peer configurations dominate anti-aligned ones conditional on the magnitude, yielding one-step stochastic dominance; composing the bucket-correction path proves the theorem.  As bucket orders generalize top-set reports, we have the following corollary.   
\begin{corollary}[Top-set elicitation]\label{cor:topset}
For any strict Bayesian SOC information structure and fixed top-set sizes $(K_n)_{n\in[N]}$, the Kendall tau mechanism is stochastically dominant truthful for top-$K_n$ reports.
\end{corollary}

In particular, for top sets $r,r'\subseteq[M]$ with comparison matrices $\mD,\mD'$, the specialized score has the overlap identity
\[
\sum_{i<j}\mD(i,j)\mD'(i,j)=M|r\cap r'|-|r||r'|.
\]
Thus, for fixed reported set sizes, the mechanism rewards overlap with peers' top sets.
Equivalently, if a top set is encoded as a binary inclusion vector, this score is an affine transformation of the Enforced Agreement (EA) score~\cite{zhang2025stochastically} on binary reports as discussed in~\cref{app:topset}.

\paragraph{More general partial-order targets.}
The proof of \cref{thm:bucket} uses two features of bucket orders: each correcting step is a transposition $\sigma_{ij}$ that relabels one report into another, and the resulting comparison-matrix increment is aligned with the corrected pair. More general partial-order targets can create dependencies among elicited comparisons, so extending stochastic-dominance truthfulness beyond fixed-shape bucket orders may require different arguments.

\section{Experiments}\label{sec:exp}

\subsection{Agent Payments Comparison}
\label{sec.ecdf}
\begin{figure*}[t]
    \centering
    \begin{subfigure}[t]{0.32\linewidth}
        \centering
        \includegraphics[width=\linewidth]{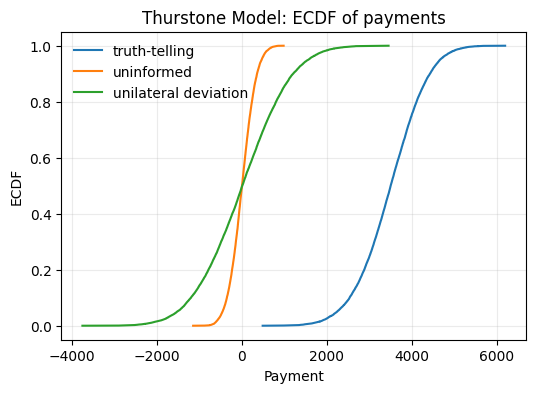}
        \caption{Thurstone (synthetic)}
        \label{fig:ecdf_thurstone}
    \end{subfigure}\hfill
    \begin{subfigure}[t]{0.32\linewidth}
        \centering
        \includegraphics[width=\linewidth]{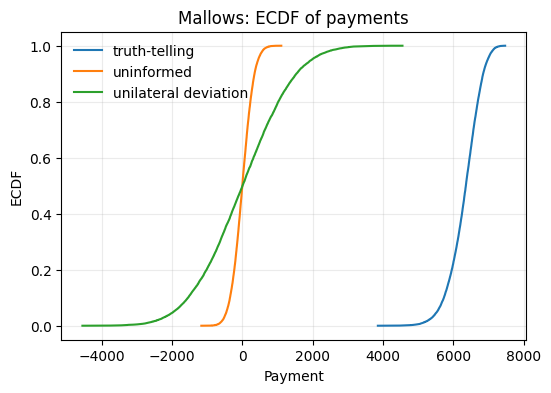}
        \caption{Mallows (synthetic)}
        \label{fig:ecdf_mallows}
    \end{subfigure}\hfill
    \begin{subfigure}[t]{0.32\linewidth}
        \centering
        \includegraphics[width=\linewidth]{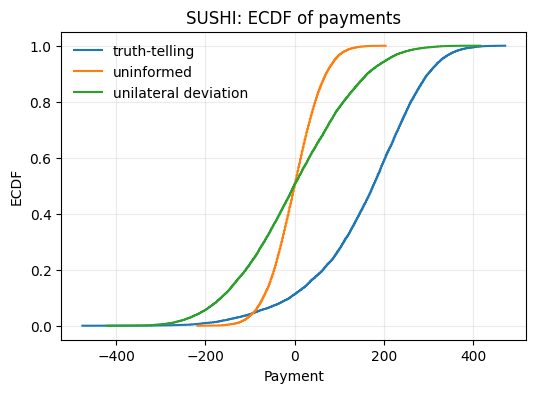}
        \caption{SUSHI (real-world)}
        \label{fig:ecdf_sushi}
    \end{subfigure}
    \caption{ECDFs of agent payments under three reporting regimes: truth-telling (observed human rankings for SUSHI), uninformed reporting (uniform random permutations), and unilateral deviation (one agent reports a random permutation while others report truthfully or by their observed SUSHI rankings).}
    \label{fig:ecdf_all}
\end{figure*}

\Cref{fig:ecdf_all} compares the empirical payment distributions of the Kendall tau peer-prediction mechanism under truthful reporting, uninformed random reporting, and unilateral random deviation. 

In each trial, agents submit full rankings and payments are computed using the score in~\cref{eq:score_full}. For the synthetic experiments (\cref{fig:ecdf_thurstone,fig:ecdf_mallows}), we use $N=25$ agents, $M=30$ items, and $600$ independent trials. In the Thurstone experiment (\cref{fig:ecdf_thurstone}), item costs are drawn i.i.d.\ from a standard Gaussian distribution, and each agent observes the costs perturbed by Gaussian noise with variance $\eta_n = 1$, as in~\cref{ex:para}. In the Mallows experiment (\cref{fig:ecdf_mallows}), a latent reference ranking is sampled and agents' truthful rankings are noisy Mallows perturbations of this reference with dispersion parameter $\eta_n=0.8$ as defined in~\cref{ex:mallows}. For SUSHI~\cite{kamishima_sushi_dataset} (\cref{fig:ecdf_sushi}), we treat the observed human ranking as the truthful/observed report: each trial samples $N=25$ observed human rankings over $M=10$ sushi items. This real-data setting is meant as an empirical stress test rather than a model guaranteed to satisfy Bayesian SOC.

For each model or dataset, we compare three reporting regimes. 
\begin{enumerate}
    \item In \emph{truth-telling}, every agent reports the ranking induced by their signal. For SUSHI, this means the observed human ranking. 
    \item In \emph{uninformed reporting}, every report is an independent uniformly random ranking. 
    \item In \emph{unilateral deviation}, one focal agent reports a uniformly random ranking while peers report truthfully or by their observed SUSHI rankings. Repeating over focal agents and trials gives the deviation distribution.
\end{enumerate}
For each regime, we pool payments across trials and plot their empirical cumulative distribution function (ECDF), so a lower, right-shifted curve corresponds to higher payments. Thus stochastic dominance of truthful reporting (\cref{def:sdt}) means the truthful ECDF lies weakly below the deviation ECDF at every threshold.

Across the synthetic Bayesian SOC models (Thurstone and Mallows), the empirical payment distributions closely match the theoretical prediction of stochastically dominant truthfulness, with the ECDF of truthful reporting lying uniformly below, or equivalently right-shifted relative to, the alternatives in our simulations (\cref{fig:ecdf_thurstone,fig:ecdf_mallows}).
On the real-world SUSHI dataset, while the dominance does not hold everywhere, the observed-report regime nevertheless has substantially more mass on higher payments (\cref{fig:ecdf_sushi}).
Taken together, these results support the distributional incentive prediction in synthetic SOC models and show favorable observed-report score distributions on heterogeneous, nonparametric preference data. In~\cref{app:structured_deviations}, we repeat the comparison against structured unilateral deviations, including near-truthful perturbations and coordination-style reports; the synthetic results remain favorable to truthful reporting, and the SUSHI observed-report distribution is generally the most favorable in this real-data stress test.

\subsection{Sensitivity to Report Quality}\label{sec:sensitivity}
In~\cref{sec.ecdf}, we test the distributional incentive prediction. We now ask whether the payment is responsive to the quality of an agent's report. We measure this by normalized sensitivity in the synthetic models and by a report-degradation test on SUSHI.

\paragraph{Synthetic models.}
In the Thurstone and Mallows models, we designate one focal agent while all other agents report truthfully, and we vary the focal agent's noise or dispersion parameter $\eta_n$, where a larger $\eta_n$ corresponds to noisier signals or lower effective effort. Let $U_n(\eta_n)$ denote the focal agent's payment. Following~\cite{ZhangS2023Higheffort}, we measure normalized sensitivity by
\[
\mathrm{Sens}_n
=
\frac{- \frac{d}{d\eta_n}\mathbb{E}[U_n(\eta_n)]}{\sigma[U_n(\eta_n)]},
\]
so larger sensitivity values mean that expected payment falls more sharply, relative to payoff variability, as the focal agent's signal becomes noisier.

We fix $N=25$ agents and sweep the number of items $M$. In the Thurstone model, item costs $\vtheta \in \mathbb{R}^M$ are drawn i.i.d.\ from $\mathcal{N}(0,1)$ and each agent observes them corrupted by independent Gaussian noise with variance $\eta_n$ with the baseline $\eta_n = 0.7$ for every agent. In the Mallows model, the reference ranking is drawn uniformly from $\symm_M$ and each agent's ranking is a Mallows perturbation of it with dispersion $\eta_n$ with the baseline $\eta_n = 0.35$ for every agent. Both the derivative and the standard deviation in $\mathrm{Sens}_n$ are evaluated at this baseline. We estimate the derivative by a symmetric finite difference in the focal agent's $\eta_n$, with step $\delta = 0.05$ for Thurstone and $\delta = 0.03$ for Mallows, holding all other agents at the baseline. For each setting, we use $50$ independent trials and $200$ Monte Carlo samples, with common random numbers to reduce variance.

\begin{figure*}[t]
    \centering
    \begin{subfigure}[t]{0.32\linewidth}
        \centering
        \includegraphics[width=\linewidth]{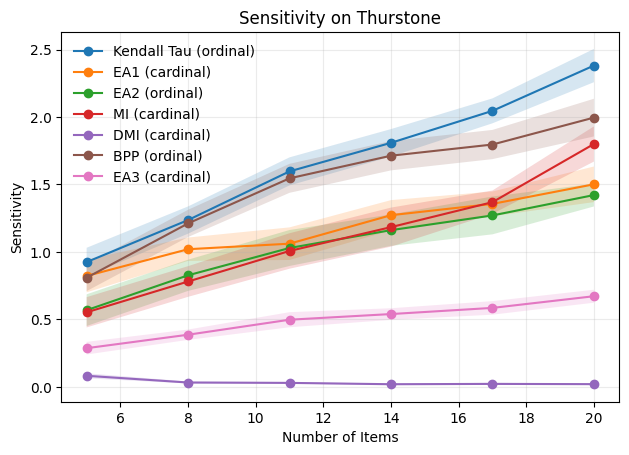}
        \caption{Thurstone sensitivity}
        \label{fig:sens_thurstone}
    \end{subfigure}\hfill
    \begin{subfigure}[t]{0.32\linewidth}
        \centering
        \includegraphics[width=\linewidth]{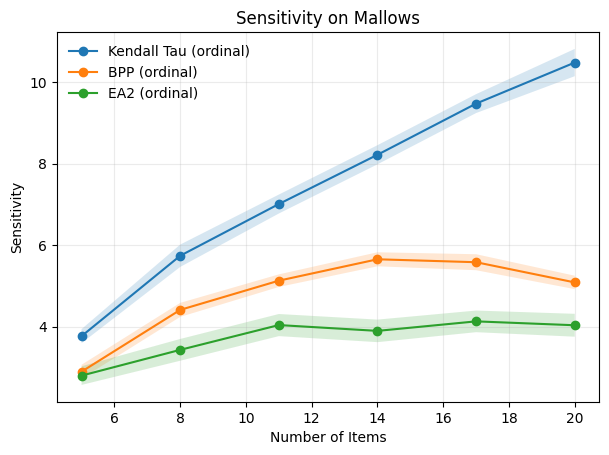}
        \caption{Mallows sensitivity}
        \label{fig:sens_mallows}
    \end{subfigure}\hfill
    \begin{subfigure}[t]{0.32\linewidth}
        \centering
        \includegraphics[width=\linewidth]{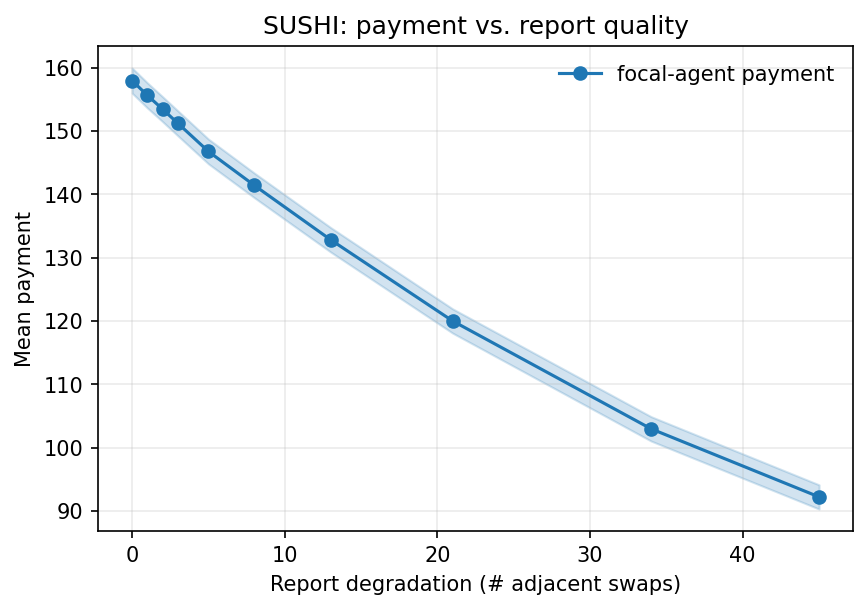}
        \caption{SUSHI degradation}
        \label{fig:sens_sushi}
    \end{subfigure}
    \caption{Measurement-quality tests. Panels~\subref{fig:sens_thurstone}--\subref{fig:sens_mallows} compare normalized sensitivity for Kendall tau and baseline mechanisms under the Thurstone and Mallows models. Panel~\subref{fig:sens_sushi} degrades a focal observed SUSHI ranking by random adjacent swaps and plots mean focal-agent payment with a 95\% confidence band.}
    \label{fig:sensitivity}
\end{figure*}

We compare Kendall tau scoring with six baselines.  In \cref{fig:sensitivity}, the cardinal baselines observe the agents' numerical signals: EA1 and EA3 apply Enforced Agreement~\cite{zhang2025stochastically} after binarizing each signal by its sign or by whether it exceeds the average signal on that item, respectively. MI~\cite{kong2019information} scores empirical Shannon mutual information after discretizing signals into three quantile buckets, and DMI uses determinant mutual information~\cite{kong2020dominantly}. The ordinal baselines observe only rankings: EA2 applies Enforced Agreement to each ranking's top-half inclusion vector, and BPP~\cite{chen2024carrot} treats sampled unordered item pairs as independent binary comparison tasks.

\Cref{fig:sens_thurstone} compares against all six baselines. \Cref{fig:sens_mallows} compares only against the ordinal baselines EA2 and BPP, since the Mallows model produces rankings rather than numerical signals. In both models, the Kendall tau mechanism has the largest normalized sensitivity among the tested mechanisms as the number of items grows. Since its sensitivity is positive, a small unilateral increase in the focal agent's noise or dispersion locally lowers that agent's expected payment.

\paragraph{SUSHI degradation.}
SUSHI has no observed signal-noise or effort parameter, so we instead degrade a focal agent's observed ranking directly. Using the sampling of~\cref{sec.ecdf} ($N=25$ observed rankings over $M=10$ items per trial), we apply $k\in\{0,1,2,3,5,8,13,21,34,45\}$ random adjacent transpositions to the focal agent's ranking while all peers keep their observed rankings, and record the focal Kendall tau payment. Here $k=0$ is the observed ranking and $k=45=\binom{10}{2}$ approaches a random report. We average the focal payment over all $N$ focal agents and $600$ trials; the band in~\cref{fig:sens_sushi} is a normal-approximation 95\% confidence interval for this mean. Mean payment decreases as the observed report is progressively degraded (\cref{fig:sens_sushi}).

Thus, in these settings, the stochastic-dominance guarantee does not appear to reduce local sensitivity, and the score also tracks report quality on real preference data.

\section{Discussion and Future Work}\label{sec:conclusion}
We introduce a simple peer-prediction mechanism that rewards pairwise agreement using Kendall tau, giving a general framework for ordinal information elicitation. Under Bayesian Stochastically Order Consistent (SOC) information structures---a broad class including Mallows and Thurstone models---truthful reporting stochastically dominates any deviation. Unlike mechanisms that rely on learning or strong prior assumptions, our mechanism handles heterogeneous agents with complex, continuous signal spaces. It also elicits fixed-shape bucket orders, including top-$K$ sets, and handles subset assignments. Experiments on synthetic data and SUSHI preferences suggest that truthful or observed reporting receives more favorable score distributions than uninformed or deviating strategies in the tested settings.

While our mechanism has these advantages, there are also limitations. The largest is our assumption of Stochastically Order Consistent (Bayesian SOC), which is satisfied by standard ranking distributions but might not always hold. One example is the case of ``cheap signals''~\cite{gao2016incentivizing,KongS2018expert}. Here agents have additional information on which they can coordinate besides the quality of the items that they are ranking. For example, instead of ranking peer reviews according to their overall quality, agents could rank them according to the institutions of the authors. This is a known issue in the peer-review literature, and our mechanism would not work as well in these cases.  Our index-equivariance assumption in~\cref{def:equivariant} excludes this setting.

Another limitation occurs when not every agent evaluates every item. We can only reward an ordering of items $i$ and $j$ if another agent also orders them. This still allows flexible assignments---not every agent needs every item---but other mechanisms allow even more general assignments~\cite{schoenebeck2020learning,chen2024carrot}.

Future research may investigate generalizations of the Bayesian SOC assumption, exploring whether our mechanism extends to alternative conditions such as Multivariate Total Positivity of order 2~\cite{fallat2016totalpositivitymarkovstructures}.

%% file: sections/appendix.tex
\section*{Appendix}
\section{Preliminaries on Partial Orderings}\label{app:pre}
A note on notation: as in the main text, we write probabilities and sums for readability. In continuous state or signal spaces, the corresponding statements are interpreted using densities and integrals with respect to common dominating measures.

A (strict) \emph{ordered set} $\mathcal{Y}$ has a strict total order so that for all $y, y',y''\in \mathcal{Y}$
\begin{enumerate}
    \item If $y<y'$, then not $y'<y$ (asymmetric)
    \item If $y<y'$, $y'<y''$, then $y<y''$ (transitive)
    \item If $y\neq y'$, then $y<y'$ or $y'<y$ (complete)
\end{enumerate}
A (strict) \emph{partially ordered set} relaxes completeness.

 \section{Full-Ranking Proofs and Details for~\cref{thm:sd}}\label{app:full}

 \begin{proof}[Proof of \cref{lem:consist}]
     Since the conditioning signal has positive marginal mass or density, it suffices to show that $\Pr[\rvx' = \vx', \rvx = \vx]>\Pr[\rvx' = \sigma(\vx'), \rvx = \vx]$. The swapped joint likelihood may be zero. By the definition of a strict Bayesian SOC model, there is a prior $\mu$ and conditional probabilities $P$ and $P'$ such that
     \begin{align*}
         &\Pr[\rvx' = \vx', \rvx =  \vx]\\
         =& \sum_{\vtheta} P'(\vx'| \vtheta)P(\vx| \vtheta)\mu(\vtheta)\\
         =& \frac{1}{2}\left(\sum_{\vtheta} P'(\vx'| \vtheta)P(\vx| \vtheta)\mu(\vtheta)+\sum_{\vtheta} P'(\vx'| \sigma(\vtheta))P(\vx| \sigma(\vtheta))\mu(\sigma(\vtheta))\right)\tag{add twice}\\
         =& \frac{1}{2}\left(\sum_{\vtheta} P'(\vx'| \vtheta)P(\vx| \vtheta)\mu(\vtheta)+\sum_{\vtheta} P'(\vx'| \sigma(\vtheta))P(\vx| \sigma(\vtheta))\mu(\vtheta)\right)\tag{$\mu$ exchangeable}\\
         =& \frac{1}{2}\sum_{\vtheta}\left( P'(\vx'| \vtheta)P(\vx| \vtheta)+P'(\sigma(\vx')| \vtheta)P(\sigma(\vx)| \vtheta)\right)\mu(\vtheta)\tag{neutrality and $\sigma(\sigma(\vtheta)) = \vtheta$}
     \end{align*}
     Therefore, the difference is
\begin{align*}
         &\Pr[\rvx' = \vx', \rvx =  \vx]-\Pr[\rvx' = \sigma(\vx'), \rvx =  \vx]\\
         =&\frac{1}{2}\sum_{\vtheta}\big( P'(\vx'| \vtheta)P(\vx| \vtheta)+P'(\sigma(\vx')| \vtheta)P(\sigma(\vx)| \vtheta)\\
         &-P'(\sigma(\vx')| \vtheta)P(\vx| \vtheta)-P'(\vx'| \vtheta)P(\sigma(\vx)| \vtheta)\big)\mu(\vtheta)\\
         =& \frac{1}{2}\sum_{\vtheta}\left( P'(\vx'| \vtheta)-P'(\sigma(\vx')| \vtheta)\right)\left( P(\vx| \vtheta)-P(\sigma(\vx)| \vtheta)\right)\mu(\vtheta)
     \end{align*}
     which depends on 
     \begin{equation}
         \left( P'(\vx'| \vtheta)-P'(\sigma(\vx')| \vtheta)\right)\left( P(\vx| \vtheta)-P(\sigma(\vx)| \vtheta)\right).\label{eq:consistent1}
     \end{equation}
     If $\vtheta$ is order consistent with $\vx$ at $(i,j)$, and thus consistent with $\vx'$, then by \cref{def:soc}
     $P'(\vx'| \vtheta)>P'(\sigma(\vx')| \vtheta)$ and $P(\vx| \vtheta)>P(\sigma(\vx)| \vtheta)$, so \cref{eq:consistent1} is positive. Similarly, if $\vtheta$ is order consistent with $\sigma(\vx)$, \cref{eq:consistent1} is also positive. Therefore, $\Pr[\rvx' = \vx', \rvx =  \vx]>\Pr[\rvx' = \sigma(\vx'), \rvx =  \vx]$, which completes the proof.
 \end{proof}

\begin{proof}[Proof of \cref{thm:truthful}]
By~\cref{def:comp}, the mechanism depends only on pairwise comparisons. Hence, we can extend the Kendall tau mechanism to take comparison matrices as inputs rather than rankings. We show that this extended mechanism maintains truthfulness even when agents report arbitrary comparison matrices that do not correspond to any valid ranking.

Consider agent $n$. If all other agents report truthfully, we need to show that truth-telling maximizes the agent's score. By~\cref{eq:kendall_comp}, the agent's best response consists of choosing $\mD(i,j)$ for each pair $(i,j)$ separately to maximize her score. For any $(i,j)$ with $x_{n,i}<x_{n,j}$, the score of truth-telling on pair $(i,j)$ with agent $n'$ is
$$\E[\mD_{n'}(i,j)\mid \vx_{n}] = \Pr[\rx_{n',i}<\rx_{n',j}\mid \vx_n]-\Pr[\rx_{n',i}>\rx_{n',j}\mid \vx_n]>0$$
by~\cref{lem:consist}, and it is larger than the score of reporting the wrong comparison $-\E[\mD_{n'}(i,j)\mid \vx_{n}]$. The case $x_i>x_j$ holds by symmetry. Summing over all peers $n'$ and pairs $(i,j)$, the unique best response is the comparison matrix consistent with the signal $\vx_{n}$, corresponding to the truthful ranking $ord(\vx_{n}) = \Gamma_n(\vx_n)$.
\end{proof}

\begin{proof}[Proof of \cref{lem:pair1}]
We can decompose the score into three parts.
\begin{align*}
    &U^t(\rho, \va)
    = U^t(\mX_{-n})\\
    =& \sum_{n'}\Bigg(
        \sum_{\substack{i<j\\i,j\notin \{i_t, j_t\}}}\sign(x_{n', i},x_{n', j};r^t_i, r^t_j)+
        \sum_{\substack{i<j\\|\{i, j\}\cap \{i_t, j_t\}| = 1}}
        \sign(x_{n', i},x_{n', j};r^t_i, r^t_j)
    \Bigg)+L(\va),
\end{align*}
because
\[
L(\va)=\sum_{n'} \sign(x_{n',i_t}, x_{n',j_t};x_{n,i_t}, x_{n,j_t}),
\]
and
\[
\sign(x_{n',i_t}, x_{n',j_t};x_{n,i_t}, x_{n,j_t})
= \sign(x_{n',i_t}, x_{n',j_t};r^t_{i_t}, r^t_{j_t}).
\]
For the difference $U^t(\rho,\va)-U^t(\rho,\va')$, the first terms agree because of the same class.  Hence
\begin{align*}
    &U^t(\rho,\va)-U^t(\rho,\va') =  U^t(\mX_{-n})-U^t(\mX_{-n}')\\
    =& \sum_{n'}\sum_{\substack{i<j\\|\{i, j\}\cap \{i_t, j_t\}| = 1}}
    \Big(\sign(x_{n', i},x_{n', j};r^t_i, r^t_j)\\
    &\qquad\qquad\qquad
    -\sign(x_{n', i}',x_{n', j}';r^t_i, r^t_j)\Big)
    +L(\va)-L(\va')
\end{align*}
Finally, each term with $n'$ and $j\notin \{i_t, j_t\}$ is equal to zero, because
\begin{align*}
    &\sum_{i\in\{i_t, j_t\}}\left(\sign(x_{n', i},x_{n', j};r^t_i, r^t_j)-\sign(x_{n', i}',x_{n', j}';r^t_i, r^t_j)\right)\\
    =& \sum_{i\in\{i_t, j_t\}}\left(\sign(x_{n', i},x_{n', j})-\sign(x_{n', i}',x_{n', j}')\right)\sign(r^t_i, r^t_j)\\
    =& \left(\sign(x_{n', i_t},x_{n', j})-\sign(x_{n', i_t}',x_{n', j}')\right)\sign(k_t, r^t_j)\\
    &+\left(\sign(x_{n', j_t},x_{n', j})-\sign(x_{n', j_t}',x_{n', j}')\right)\sign(k_t+1, r^t_j)\\
    =& \left(\sign(x_{n', i_t},x_{n', j})-\sign(x_{n', i_t}',x_{n', j}')+\sign(x_{n', j_t},x_{n', j})-\sign(x_{n', j_t}',x_{n', j}')\right)\\
    &\cdot \sign(k_t, r^t_j)\tag{because $\sign(k_t+1, r^t_j) = \sign(k_t, r^t_j)$}\\
    =&0\tag{$\{x_{n', i_t}, x_{n', j_t}\} = \{x_{n', i_t}', x_{n', j_t}'\}$}
\end{align*}
This completes the proof of \cref{eq:sd_pair11}.

Finally, swapping the $(i_t, j_t)$ pair only affects the score difference between $r^t$ and $r^{t-1}$ on pair $(i_t, j_t)$ across the $N-1$ peers. Hence
$U^{t-1}(\rho, \va) = U^t(\rho,\va)-2L(\va)$, which proves \cref{eq:sd_pair12}.
\end{proof}

\begin{proof}[Proof of \cref{lem:pair2}]
Fix $\vx_n$, $\rho$, and a correcting pair $(i,j)=(i_t,j_t)$ with $x_{n,i}<x_{n,j}$. Let $\sigma$ swap this pair, and write $m=N-1$. If $d\notin\{-m,-m+2,\dots,m\}$, both sides are zero.

For each state $\vtheta$ and peer $n'\ne n$, let $p_{n',\vtheta}(1)=P_{n'}(\hat{\vx}_{n'}\mid\vtheta)$ and $p_{n',\vtheta}(-1)=P_{n'}(\check{\vx}_{n'}\mid\vtheta)$ be the likelihoods of the two orientations at the fixed magnitude. Conditional independence gives the likelihood of magnitude $\rho$ and margin $d$ in state $\vtheta$ as
\[
    W_{\vtheta}(d):=\sum_{\va:L(\va)=d}\prod_{n'\ne n}p_{n',\vtheta}(a_{n'}).
\]
Neutrality implies $W_{\sigma(\vtheta)}(d)=W_{\vtheta}(-d)$. Write $\pi_n(\vx_n)$ for the positive marginal mass or density of the focal signal. Pairing each state with its swap using exchangeability, and using the absence of state ties, gives
\begin{equation}\label{eq:sd_pair4}
\begin{aligned}
    &\pi_n(\vx_n)\bigl(Q(\rho,L=d)-Q(\rho,L=-d)\bigr)\\
    &\quad=\sum_{\vtheta:\theta_i<\theta_j}
        \mu(\vtheta)\bigl(P_n(\vx_n\mid\vtheta)-P_n(\vx_n\mid\sigma(\vtheta))\bigr)\\
    &\qquad\qquad\cdot\bigl(W_{\vtheta}(d)-W_{\vtheta}(-d)\bigr).
\end{aligned}
\end{equation}
In continuous models the sum is an integral with respect to the prior; the likelihoods are densities under the convention of~\cref{sec:pre_rank}.

For each state in this sum, SOC and neutrality give $P_n(\vx_n\mid\vtheta)\ge P_n(\vx_n\mid\sigma(\vtheta))$ and $p_{n',\vtheta}(1)\ge p_{n',\vtheta}(-1)$ for every peer. To compare the margin likelihoods even when some weights are zero, add $\epsilon>0$ to both orientation weights of every peer. Denote the resulting product-sum by $W_{\vtheta,\epsilon}(d)$. These are auxiliary weights, not a new probability model. Let $k=(m+d)/2$, and set
\[
    A_\epsilon=\prod_{n'\ne n}\bigl(p_{n',\vtheta}(-1)+\epsilon\bigr),
    \qquad
    \lambda_{n'}=\frac{p_{n',\vtheta}(1)+\epsilon}{p_{n',\vtheta}(-1)+\epsilon}\ge 1.
\]
An alignment with margin $d$ has exactly $k$ positive signs. Thus
\[
    W_{\vtheta,\epsilon}(d)-W_{\vtheta,\epsilon}(-d)
    = A_\epsilon\bigl(\mathcal E_k(\bm\lambda)-\mathcal E_{m-k}(\bm\lambda)\bigr)
    \ge 0
\]
by~\cref{lem:symm_poly}. Each product-sum is a polynomial in the weights, so letting $\epsilon\downarrow 0$ at this fixed state gives $W_{\vtheta}(d)\ge W_{\vtheta}(-d)$. All summands in~\cref{eq:sd_pair4} are therefore nonnegative, proving~\cref{eq:sd_pair3}.
\end{proof}

For $\vz=(z_1,\dots,z_m)$, the elementary symmetric polynomial of degree $k$ is
\begin{equation}\label{def:symm_poly}
    \mathcal E_k(\vz)=\sum_{S\in\binom{[m]}{k}}\prod_{\ell\in S}z_\ell,
    \qquad \mathcal E_0(\vz)=1.
\end{equation}
\begin{lemma}\label{lem:symm_poly}
If $z_\ell\ge 1$ for all $\ell$ and $k$ is an integer with $m/2<k\le m$, then $\mathcal E_k(\vz)\ge\mathcal E_{m-k}(\vz)$. If all $z_\ell>1$, the inequality is strict.
\end{lemma}

\begin{proof}
Take a uniformly random permutation of $[m]$. Let $S$ contain its first $m-k$ indices and $T$ its first $k$ indices. Then $S\subset T$, and both sets are uniform among sets of their respective sizes. Since $z_\ell\ge1$, the product over $T$ is at least the product over $S$ for every permutation. Therefore
\begin{equation}\label{eq:symm_poly1}
    \frac{\mathcal E_k(\vz)}{\binom{m}{k}}
    =\E\Bigl[\prod_{\ell\in T}z_\ell\Bigr]
    \ge \E\Bigl[\prod_{\ell\in S}z_\ell\Bigr]
    =\frac{\mathcal E_{m-k}(\vz)}{\binom{m}{m-k}}.
\end{equation}
The two denominators agree. When every $z_\ell>1$, the product comparison is strict because $T\setminus S$ is nonempty.
\end{proof}

\begin{remark}
Without full support, both likelihoods of a feasible margin and its negation can be zero. The one-step dominance proof uses only the weak comparison in~\cref{lem:pair2}; strictness of the final incentive guarantee follows separately from a strict expected-score gain.
\end{remark}

\section{Proofs and Details for \Cref{thm:strong}}\label{app:strong}

\begin{proof}
Fix an index-equivariant profile $\valpha$, and as in \cref{eq:kendall_comp}
let $\tilde{\mD}_n = comp(\alpha_n(\rvx_n))$ denote agent $n$'s random
reported comparison matrix. By linearity,
\begin{equation}\label{eq:decomp}
\E\left[U_n(\valpha(\rmX))\right]
= \sum_{n' \ne n} \sum_{i<j}
\E\left[ \tilde{\mD}_n(i,j) \tilde{\mD}_{n'}(i,j) \right],
\end{equation}
so we can focus on one peer $n' \ne n$ and one item pair $(i,j)$ with
$i < j$. Let $\sigma = \sigma_{ij}$.

We define the \emph{alignment factor} of strategy $\alpha_n$ at $(i,j)$,
\begin{equation*}
\kappa_n(\vx) := 1 - 2\Pr\left[ \tilde{\mD}_n(i,j) \neq \sign(x_i, x_j)
 \mid \rvx_n = \vx \right] \in [-1,1].
\end{equation*}  We can write the expected reported
comparison as
$\E\left[ \tilde{\mD}_n(i,j) \mid \rvx_n = \vx \right]
= \kappa_n(\vx) \sign(x_i, x_j)$, with $\kappa_n \equiv 1$ for
truth-telling. Moreover, $\kappa_n(\sigma(\vx)) = \kappa_n(\vx)$ because
the strategy is index-equivariant. Since the strategies'
randomness is independent across agents and of all signals, conditioning
on $(\rvx_n, \rvx_{n'})$ gives
\begin{equation*}
\E\left[ \tilde{\mD}_n(i,j) \tilde{\mD}_{n'}(i,j) \right]
= \E\left[ \kappa_n(\rvx_n) \kappa_{n'}(\rvx_{n'})
\sign(\rx_{n,i}, \rx_{n,j};\ \rx_{n',i}, \rx_{n',j}) \right].
\end{equation*}
 
In the first part, we prove
\begin{equation}\label{eq:pairbound}
\E\left[ \tilde{\mD}_n(i,j) \tilde{\mD}_{n'}(i,j) \right]
\le
\E\left[ \sign(\rx_{n,i}, \rx_{n,j};\ \rx_{n',i}, \rx_{n',j}) \right],
\end{equation}
whose right side is the same pair term under truth-telling. Partition the
pairs of signal vectors into groups of four,
$\{(\vx, \vx'), (\sigma(\vx), \vx'), (\vx, \sigma(\vx')),
(\sigma(\vx), \sigma(\vx'))\}$ with the representative satisfying
$\sign(x_i, x_j) = \sign(x'_i, x'_j) = 1$, and let
$P(\vx,\vx') := \Pr[\rvx_n = \vx, \rvx_{n'} = \vx']$. The contribution of
one group to the left side of \cref{eq:pairbound} is
\begin{align*}
    &\kappa_n(\vx)\kappa_{n'}(\vx')P(\vx, \vx')
    - \kappa_n(\sigma(\vx))\kappa_{n'}(\vx')P(\sigma(\vx), \vx')\\
    &\quad - \kappa_n(\vx)\kappa_{n'}(\sigma(\vx'))P(\vx, \sigma(\vx'))
    + \kappa_n(\sigma(\vx))\kappa_{n'}(\sigma(\vx'))P(\sigma(\vx), \sigma(\vx'))\\
    &= \kappa_n(\vx)\kappa_{n'}(\vx')
    \BigParens{P(\vx, \vx') - P(\sigma(\vx), \vx') - P(\vx, \sigma(\vx'))
    + P(\sigma(\vx), \sigma(\vx'))}
    \tag{$\kappa_n(\sigma(\vx)) = \kappa_n(\vx)$}\\
    &\le P(\vx, \vx') - P(\sigma(\vx), \vx') - P(\vx, \sigma(\vx'))
    + P(\sigma(\vx), \sigma(\vx')),
\end{align*}
where the inequality uses that $|\kappa_n(\vx)\kappa_{n'}(\vx')| \le 1$ and the group's truthful contribution is
positive,
\[
\BigParens{P(\vx,\vx') - P(\vx, \sigma(\vx'))}
+ \BigParens{P(\sigma(\vx), \sigma(\vx')) - P(\sigma(\vx), \vx')} > 0,
\]
by \cref{lem:consist}. Summing over groups yields
\cref{eq:pairbound}, with equality if and only if
$\kappa_n(\vx) \kappa_{n'}(\vx') = 1$ on every group of positive
probability. Summing \cref{eq:pairbound} over item pairs
and peers proves
\[
\E[U_n(\valpha(\rmX))] \le \E[U_n(\Gamma(\rmX))].
\]

For the second part, when the agent $n$ gets the highest expected score $\E[U_n(\valpha(\rmX))] =  \E[U_n(\Gamma(\rmX))]$, equality holds in
\cref{eq:pairbound} for every peer $n'$ and every item pair $(i,j)$. By
full support, $\kappa_n(\vx) \kappa_{n'}(\vx') = 1$ for almost all $(\vx, \vx')$. Hence $|\kappa_n| = |\kappa_{n'}| = 1$, and fixing any
$\vx'$ shows $\kappa_n(\vx) = \kappa_{n'}(\vx') = \epsilon^{ij}
\in \{\pm 1\}$, independent of the signal and common to all agents.   
Finally, by index-equivariance, the alignment factors agree across item pairs, 
$\tilde{\mD}_n(i,j) = \epsilon \sign(\rx_{n,i}, \rx_{n,j})$ almost surely
for every pair. If $\epsilon = 1$, every agent reports $\Gamma_n(\vx_n)$.  If $\epsilon = -1$, every agent reports the reversed ranking
$rev \circ \Gamma_n(\vx_n)$.
\end{proof}

\section{Subset-Assignment Proofs and Details in~\cref{sec:subset}}\label{app:subset}

The proof has two steps. \Cref{lem:subset-step} follows the correction path of a misreport on $S_n$ and shows that, for a fixed assignment and peer, each step that swaps two shared items strictly improves the score distribution, while every other step leaves the score unchanged. The proof of~\cref{thm:subset} then averages each step over assignments and peers and composes the steps. Let agent $n$'s score at step $t$ of the correction path against $n'$ be
\begin{equation*}
    U^t_{n'}:=\sum_{i,j\in S_{n}\cap S_{n'}: i<j}\mD_{r^t}(i,j)\,\mD_{r_{n'}}(i,j),
\end{equation*}
where $r_{n'}=\Gamma_{n'}(\vx_{n'})$.  

\begin{lemma}[One correcting step]\label{lem:subset-step}
Fix agent $n$'s signal $\vx_n$, a report $\tilde r$ with correction path $r^0,\dots,r^T$ on $S_n$, a realization of the assignment $\gS$, and a drawn peer $n'$. Then for every $t=1,\dots,T$, conditional on $(\vx_n,\gS,n')$,
\begin{equation*}
    \begin{cases}
        U^t_{n'}>_{\mathrm{FOSD}}U^{t-1}_{n'} & \text{if } i_t,j_t\in S_n\cap S_{n'},\\
        U^t_{n'}=U^{t-1}_{n'} & \text{otherwise.}
    \end{cases}
\end{equation*}
\end{lemma}

\begin{proof}
Write $U^s(\vx)$ for the value of $U^s_{n'}$ when the peer's signal is $\vx_{n'}=\vx$. Because $i_t$ and $j_t$ are adjacent in $r^{t-1}$, if the pair is not contained in $S_n\cap S_{n'}$, swapping it leaves the restricted ranking unchanged and hence $U^t(\vx)=U^{t-1}(\vx)$ for every $\vx$.

Now suppose $i_t,j_t\in S_n\cap S_{n'}$. They are still adjacent on the shared set, so the identities of~\cref{lem:pair1} apply to the score on that set:
\begin{align*}
    U^t_{n'}(\rho,a_{n'})-U^t_{n'}(\rho,-a_{n'})&=2a_{n'},\\
    U^t_{n'}(\rho,a_{n'})-U^{t-1}_{n'}(\rho,a_{n'})&=2a_{n'}.
\end{align*}
Fixing the assignment and sampled peer does not change the law of that peer's signal given $\vx_n$. Apply~\cref{lem:pair2} to the two-agent marginal containing $n$ and $n'$, with $L=a_{n'}$. Writing $u(\rho)=U^t_{n'}(\rho,1)$, we obtain for every $w$
\begin{align*}
    &\Pr[U^t_{n'}\ge w\mid\vx_n,\gS,n']-\Pr[U^{t-1}_{n'}\ge w\mid\vx_n,\gS,n']\\
    &\quad=\sum_{\rho}\bigl(\vone[u(\rho)\ge w]-\vone[u(\rho)-2\ge w]\bigr)\\
    &\qquad\qquad\cdot\bigl(Q(\rho,L=1)-Q(\rho,L=-1)\bigr)\ge 0.
\end{align*}
For strictness,~\cref{lem:consist} gives $\E[a_{n'}\mid\vx_n,\gS,n']>0$, since the posterior comparison is more likely to align than to disagree. Consequently $\E[U^t_{n'}-U^{t-1}_{n'}\mid\vx_n,\gS,n']>0$. Weak dominance and this strict mean difference for bounded scores imply strict inequality at some threshold.
\end{proof}

\begin{proof}[Proof of~\cref{thm:subset}]
Fix $\vx_n$, the agent's assignment $S_n$, and a deterministic misreport different from the truthful ranking on $S_n$. The correcting path depends only on this information. By~\cref{lem:subset-step}, each step weakly improves the score distribution conditional on the complete assignment and chosen peer. Averaging over that assignment and peer conditional on $(\vx_n,S_n)$, and composing the steps, gives $U^T\ge_{\mathrm{FOSD}}U^0$.

The first correcting pair is covered with positive probability by pairwise coverage. On coverage its expected gain is strictly positive by~\cref{lem:subset-step}, and otherwise the gain is zero. Later steps have nonnegative expected gains. Hence $\E[U^T-U^0\mid\vx_n,S_n]>0$, so the dominance is strict at some threshold. Mixing over reports preserves weak dominance and preserves a strict mean gain for any strategy that misreports with positive probability at the given signal and assignment.
\end{proof}

\section{Bucket-Order Proofs}\label{app:bucket}
As in~\cref{sec:sec:full_sd}, we analyze the mechanism one \emph{correcting adjacent-bucket swap} at a time for agent $n$'s report.

To prove stochastic dominance, we first isolate the contribution of a single peer: for an agent report $r$ and the truthful peer report $r_{n'}=\Gamma_{n'}(\vx_{n'})$, set
$u_{n'}(r) := \langle \mD_r,\mD_{r_{n'}}\rangle$, so that $U_n(r,\mX_{-n}) = \sum_{n'\neq n} u_{n'}(r)$.

\begin{definition}[Bucket correction path]\label{def:bucketpath}
Given a shape-$\bm{K}^n$ report $\tilde r$ and the truthful target $r_n := \Gamma_n(\vx_n)$, the \emph{bucket correction path} $r^0,r^1,\dots,r^T$ is defined by $r^0=\tilde r$ and, while $r^{t-1}\neq r_n$, choosing adjacent buckets $b,b+1$, letting $i$ be the best-signal item of $B^n_{b+1}$ and $j$ the worst-signal item of $B^n_b$, and---whenever $x_{n,i}<x_{n,j}$---swapping them to form $r^t$.
\end{definition}

By transitivity of $\ge_{\mathrm{FOSD}}$ it suffices to prove that each step improves the score distribution:
\begin{equation}\label{eq:bucket_step}
U_n(r^t,\mX_{-n})\ \ge_{\mathrm{FOSD}}\ U_n(r^{t-1},\mX_{-n})\qquad\text{given }\vx_n\text{ with }x_{n,i}<x_{n,j}.
\end{equation}
With the magnitude and alignment defined in~\cref{def:events}, write $U^t(\rho,\va)=U_n(r^t,\mX_{-n})$ for the step-$t$ score under peer signals of magnitude $\rho$ and alignment $\va$.

Because $r^t=\sigma_{ij}\,r^{t-1}$, bucketing commutes with relabeling, and $\langle\cdot,\cdot\rangle$ is invariant under a common relabeling, the peer contribution satisfies
\begin{equation}\label{eq:involution}
u^t_{n'}(\sigma_{ij}\vx_{n'}) = u^{t-1}_{n'}(\vx_{n'}).
\end{equation}

The following lemma shows that we can write the difference as a linear function of the alignment vector $\va$.  

\begin{lemma}\label{lem:bucket_increment}
For any $(i_t, j_t)$, $\vx_n$, magnitude $\rho$ and alignment vectors $\va,\va'$, there is a nonnegative integer vector $\vc_t$ so that
\begin{equation}\label{eq:bucket_diff}
U^t(\rho,\va)-U^t(\rho,\va') = \tfrac12\langle\vc_t, \va-\va'\rangle
\end{equation}
and
\begin{equation}\label{eq:bucket_inc}
U^t(\rho,\va)-U^{t-1}(\rho,\va) = \langle\vc_t, \va\rangle.
\end{equation}
\end{lemma}

\begin{proof}When $i$ moves from bucket $b+1$ to $b$ and $j$ from $b$ to $b+1$, a pair's comparison can change only if it involves $i$ or $j$ and the other element lies in bucket $b$ or $b+1$: both items cross only the boundary between buckets $b$ and $b+1$, so comparisons with items in buckets $<b$ or $>b+1$ are unaffected. For the pair $(i,j)$, $\mD_{r^{t-1}}(i,j)=\sign(b+1,b)=-1$ and $\mD_{r^t}(i,j)=\sign(b,b+1)=+1$, so $\Delta\mD(i,j)=2$. For $k\in(B^n_b\cup B^n_{b+1})\setminus\{i,j\}$, with $\beta(k)\in\{b,b+1\}$,
\begin{align*}
\Delta\mD(i,k)
&=\sign(b,\beta(k))-\sign(b+1,\beta(k))=+1,\\
\Delta\mD(j,k)
&=\sign(b+1,\beta(k))-\sign(b,\beta(k))=-1,
\end{align*}
and $\Delta\mD$ vanishes on all other pairs. Thus, the difference becomes
\begin{align*}
u^t_{n'}-u^{t-1}_{n'}&= \langle \mD_{r^t},\mD_{r_{n'}}\rangle-\langle \mD_{r^{t-1}},\mD_{r_{n'}}\rangle\\
&= \langle\Delta\mD,\mD_{r_{n'}}\rangle\\
&= 2\,\mD_{r_{n'}}(i,j)
 +\!\!\!\sum_{k\in (B^n_b\cup B^n_{b+1})\setminus\{i,j\}}\!\!\!
 \big(\mD_{r_{n'}}(i,k)-\mD_{r_{n'}}(j,k)\big).
\end{align*}

Now, we want to write the difference as a function of the alignment vector, 
$u^t_{n'}-u^{t-1}_{n'} = c^t_{n'} a_{n'}.$  If $i,j$ lie in one bucket of $r_{n'}$ then $\mD_{r_{n'}}(i,j)=0$ and $\mD_{r_{n'}}(i,k)=\mD_{r_{n'}}(j,k)$ for all $k$, so $u^t_{n'}-u^{t-1}_{n'}=0$ and we can set 
$$c^t_{n'}:=0.$$
If peer $n'$ ranks $i$ strictly above $j$, i.e. $\beta_{r_{n'}}(i)<\beta_{r_{n'}}(j)$, $\mD_{r_{n'}}(i,j)=1$ contributes $2$.  For the second term, $\mD_{r_{n'}}(i,k)-\mD_{r_{n'}}(j,k)$ equals $2$ when $\beta_{r_{n'}}(i)<\beta_{r_{n'}}(k)<\beta_{r_{n'}}(j)$, $1$ when $k$ shares a bucket with exactly one of $i,j$, and $0$ otherwise, so summing the nonnegative terms gives $u^t_{n'}-u^{t-1}_{n'}=c^t_{n'}$ with
\begin{align*}
c^t_{n'}=2 + 2|\{k:\, i\prec_{n'}k\prec_{n'}j\}|+|\{k:\, k\sim_{n'}i\ \mathrm{xor}\ k\sim_{n'}j\}|
 \ge 2,
\end{align*}
the counts over $k\in(B^n_b\cup B^n_{b+1})\setminus\{i,j\}$ and $\prec_{n'},\sim_{n'}$ denoting peer $n'$'s strict bucket order and same-bucket relation, depending only on the buckets that the magnitude $\rho$ fixes. This completes the case for $a_{n'} = 1$; the case $a_{n'} = -1$ holds by symmetry. Applying $\sigma_{ij}$ to the peer's signal negates every sign above and hence the gain, so for every peer there is a coefficient $(\vc^t)_{n'}\ge0$ such that
\begin{equation}\label{eq:bucket_peer}
u^t_{n'}-u^{t-1}_{n'}=c^t_{n'}\,a_{n'}.
\end{equation}

Summing over peers gives \cref{eq:bucket_inc}, $U^t(\rho,\va)-U^{t-1}(\rho,\va)=\sum_{n'}(\vc^t)_{n'}a_{n'}=\langle\vc^t,\va\rangle$. 
For~ \cref{eq:bucket_diff}, \cref{eq:involution} gives $u^{t-1}_{n'}(\va)=u^t_{n'}(\text{flip }a_{n'})$, so $u^t_{n'}(+1)-u^t_{n'}(-1)=(\vc^t)_{n'}$; as $u^t_{n'}$ is affine in $a_{n'}\in\{-1,1\}$,
$$U^t(\rho,\va)-U^t(\rho,\va')=\sum_{n'}\big(u^t_{n'}(a_{n'})-u^t_{n'}(a'_{n'})\big)=\tfrac12\sum_{n'}(\vc^t)_{n'}\,(a_{n'}-a'_{n'})=\tfrac12\langle\vc^t,\va-\va'\rangle.$$
\end{proof}

For a full ranking ($\bm{K}=1^M$) the buckets $B^n_b,B^n_{b+1}$ are singletons, every peer splits $i,j$, and $\vc_t=2\,\vone$, so $\langle\vc_t,\va\rangle=2L(\va)$, which is \cref{lem:pair1}.

\begin{lemma}\label{lem:bucket_B}
Given $\vtheta$ with $\theta_i<\theta_j$, $U_n(r^t,\mX_{-n})\ge_{\mathrm{FOSD}}U_n(r^{t-1},\mX_{-n})$ conditional on $\vtheta$.
\end{lemma}

\begin{proof}[Proof of~\cref{lem:bucket_B}]
Conditional on $\vtheta$ the peers are independent (\cref{sec:pre_rank}).  If we condition further on the magnitude $\rho$ (\cref{def:events}), stochastic consistency (\cref{def:soc}) gives $p_{n'}:=\Pr[a_{n'}=+1\mid\vtheta,\rho]\ge\tfrac12$, the configuration $a_{n'}=+1$ being order consistent with $\vtheta$ at $(i,j)$. 

Fix a peer $n'$ and let $u_+,u_-$ be the values of $u^t_{n'}$ at $a_{n'}=+1$ and $a_{n'}=-1$; by \cref{eq:bucket_peer} the gain is $u_+-u_-=(\vc_t)_{n'}\ge0$, and by \cref{eq:involution} $u^{t-1}_{n'}$ equals $u^t_{n'}$ with $a_{n'}$ flipped, so $u^t_{n'}$ takes the larger value $u_+$ with probability $p_{n'}$ whereas $u^{t-1}_{n'}$ takes it with probability $1-p_{n'}\le p_{n'}$. As $u^t_{n'}$ and $u^{t-1}_{n'}$ both have support on $\{u_+,u_-\}$ and the former has more mass on the larger, $u^t_{n'}\ge_{\mathrm{FOSD}}u^{t-1}_{n'}$ given $(\vtheta,\rho)$. 

Finally, because stochastic dominance is preserved by sums of independent random variables~\cite{shaked2007stochastic}, summing over the independent coordinates gives
\[
U_n(r^t)
=\sum_{n'}u^t_{n'}
\ge_{\mathrm{FOSD}}
\sum_{n'}u^{t-1}_{n'}
=U_n(r^{t-1})
\]
given $(\vtheta,\rho)$; averaging over $\rho$ preserves the order.
\end{proof}

\begin{proof}[Proof of~\cref{thm:bucket}]
Fix an agent $n$, signal $\vx_n$, and misreport $\tilde r$, with correction path $r^0,\dots,r^T$ (\cref{def:bucketpath}). By transitivity of $\ge_{\mathrm{FOSD}}$, it suffices to prove the single-step dominance in~\cref{eq:bucket_step}.

Fix a step and a threshold $w\in\R$, and set
\[
G(\vtheta,w)
:=\Pr[U^t\ge w\mid\vtheta]
-\Pr[U^{t-1}\ge w\mid\vtheta],
\]
with $U^s=U_n(r^s,\mX_{-n})$. \Cref{lem:bucket_B} gives $G(\vtheta,w)\ge0$ whenever $\theta_i<\theta_j$.
Moreover, by~\cref{eq:involution}, sampling $\vx_{n'}\sim P_{n'}(\cdot\mid\sigma_{ij}\vtheta)$ is the same as sampling $\vx_{n'}\sim P_{n'}(\cdot\mid\vtheta)$ and relabeling by $\sigma_{ij}$. Hence $G(\sigma_{ij}\vtheta,w)=-G(\vtheta,w)$.

Stochastic consistency applied to the agent's own signal, together with the exchangeability of $\mu$, yields the posterior tilt $\Pr[\vtheta\mid\vx_n]\ge\Pr[\sigma_{ij}\vtheta\mid\vx_n]$ for $\theta_i<\theta_j$. Pairing each $\vtheta$ with $\sigma_{ij}\vtheta$,
\begin{align*}
&\Pr[U^t\ge w\mid\vx_n]-\Pr[U^{t-1}\ge w\mid\vx_n]\\
&\qquad =
\sum_{\vtheta:\,\theta_i<\theta_j}
\big(\Pr[\vtheta\mid\vx_n]-\Pr[\sigma_{ij}\vtheta\mid\vx_n]\big)
\,G(\vtheta,w)
\ge 0,
\end{align*}
which proves~\cref{eq:bucket_step}. Composing the $T$ steps gives
\[
U_n(r_n,\mX_{-n})
\ge_{\mathrm{FOSD}}\cdots\ge_{\mathrm{FOSD}}
U_n(r^0,\mX_{-n})
=
U_n(\tilde r,\mX_{-n}).
\]
A mixed deviation is a mixture of such reports and inherits the dominance.
\end{proof}

\cref{thm:bucket} subsumes \cref{thm:sd} (shape $1^M$) and the top-set corollary~\cref{cor:topset} (shape $(K,M-K)$). Subset assignments are treated in~\cref{app:subset}.

\section{Top-Set Score and Enforced Agreement}\label{app:topset}
Top-$K$ reports are bucket orders of shape $(K,M-K)$, so their truthfulness follows directly from~\cref{thm:bucket}. We record the specialized score identity because it explains the top-set payment and its connection to binary Enforced Agreement.

Given two partial order relations $R,R'$ with comparison matrices $\mD,\mD'\in\{-1,0,1\}^{M\times M}$ in~\cref{def:comp}, define the affine pairwise-loss functional
\begin{equation*}
    \ell_\tau(R,R')=\frac{1}{2}\sum_{i<j}(1-\mD(i,j)\mD'(i,j)).
\end{equation*}
For top-set reports $\tilde r_n,\tilde r_{n'}\subseteq[M]$ of fixed sizes, the score in~\cref{eq:score_bucket} is equivalently
\begin{equation}\label{eq:score_pord}
    U_n(\tilde{\vr})
    = \sum_{n'\neq n}\sum_{i<j}\tilde{\mD}_n(i,j)\tilde{\mD}_{n'}(i,j),
\end{equation}
where $\mD(i,j)=\vone[i\in r,j\notin r]-\vone[i\notin r,j\in r]$.

\begin{lemma}\label{lem:topset1}
    Given sets $r,r'\subseteq[M]$ with $\mD=comp(r)$ and $\mD'=comp(r')$,
    \[
        \sum_{i<j}\mD(i,j)\mD'(i,j)=M|r\cap r'|-|r||r'|.
    \]
\end{lemma}

\begin{proof}
If $i,j$ have the same membership in $r$ or the same membership in $r'$, the pair contributes zero. Agreement pairs have $i\in r\cap r'$ and $j\notin r\cup r'$, contributing $+1$, while disagreement pairs have $i\in r\setminus r'$ and $j\in r'\setminus r$, contributing $-1$. Hence
\begin{align*}
    \sum_{i<j}\mD(i,j)\mD'(i,j)
    &= |r\cap r'|(M-|r|-|r'|+|r\cap r'|)\\
    &\quad -(|r|-|r\cap r'|)(|r'|-|r\cap r'|)\\
    &= M|r\cap r'|-|r||r'|.
\end{align*}
\end{proof}

\begin{proposition}[Top-set Kendall tau and Enforced Agreement]\label{prop:kendall2ea}
For fixed set sizes $K_n,K_{n'}$ and number of items $M$, the top-set Kendall tau score is affine equivalent to the Enforced Agreement (EA) score on binary inclusion reports:
\[
    U_n(\tilde{\vr})=\frac{M}{2}\sum_{n'\neq n}S_{EA}(\tilde y_n,\tilde y_{n'})+C,
\]
where $S_{EA}(\tilde y_n,\tilde y_{n'})=\sum_{i=1}^M\vone[\tilde y_{n,i}=\tilde y_{n',i}]$ and $C$ depends on $M,K_n,K_{n'}$ but not on the reports.
\end{proposition}

\begin{proof}
For binary inclusion vectors $\tilde y_n,\tilde y_{n'}$,
\[
S_{EA}(\tilde y_n,\tilde y_{n'})
=|\tilde r_n\cap\tilde r_{n'}|+|[M]\setminus(\tilde r_n\cup\tilde r_{n'})|
=M-K_n-K_{n'}+2|\tilde r_n\cap\tilde r_{n'}|.
\]
Combining this with~\cref{lem:topset1} gives
\[
U_n(\tilde{\vr})
=\sum_{n'\neq n}\left(\frac{M}{2}S_{EA}(\tilde y_n,\tilde y_{n'})
+\frac{M}{2}(K_n+K_{n'}-M)-K_nK_{n'}\right),
\]
which proves the claim.
\end{proof}

\section{Additional Discussion of Assumptions and Related Work}\label{app:discussion}
In this section, we discuss our model assumptions and connect to the literature. 
First, we compare the Bayesian Stochastically Order Consistent (SOC) assumption with Strong Stochastic Transitivity (SST) and include a necessity observation for SOC. We then connect our framework to the probabilistic social choice literature through order independence, a relaxed form of independence of irrelevant alternatives.

\paragraph{Additional relation to prior work}
Compared to multi-task peer prediction mechanisms for categorical reports operating in discrete signal spaces, our model handles continuous underlying signals (real-valued quality scores) from which agents derive ordinal reports. This continuous setting, combined with the ordered structure of rankings, requires different analytical techniques: our proof exploits state-swap symmetry, elementary symmetric polynomial comparisons, and stochastic consistency properties rather than discrete signal correlation structures.

Our framework also bridges peer prediction and property elicitation~\cite{lambert2008eliciting}. By defining a target function that projects complex signal spaces onto structural ordinal reports (e.g., full rankings, top-sets), we may view our mechanism as a scoring rule for the ordinal shape of a belief distribution.

\paragraph{SOC and Strong Stochastic Transitivity}
To formalize this comparison, we say a conditional probability $P:\Theta^M\to \Delta(\mathcal{X}^M)$ satisfies \emph{Strong Stochastic Transitivity (SST)} if for all $i,j,k\in [M]$ such that $\Pr[\rx_i<\rx_j\mid \vtheta]>1/2$ and $\Pr[\rx_j<\rx_k\mid \vtheta]>1/2$, the following holds:
\[
    \Pr[\rx_i<\rx_k\mid \vtheta] \ge \max \{\Pr[\rx_i<\rx_j\mid \vtheta],\Pr[\rx_j<\rx_k\mid \vtheta]\}.
\]

When SOC and SST are viewed as conditions on the same full-signal distribution, SOC is stronger than SST: SOC imposes pointwise swap inequalities on full signal vectors, while SST is a pairwise condition on comparison probabilities. However, the Bayesian SOC framework and the BPP/SST framework are not directly nested as modeling frameworks. Our framework allows heterogeneous agents and richer ordinal reports such as full rankings and top sets, while the BPP mechanism of Chen et al.~\cite{chen2024carrot} is designed for homogeneous pairwise-comparison assignments. Conversely, SST does not imply SOC, because a pairwise condition need not control the likelihood of each full signal vector under item swaps.

In addition to the examples in~\cref{sec:pre_rank}, the standard random binary noise model also satisfies Bayesian SOC, though it is not strict.

\begin{example}[Binary Noise Model]\label{ex:bsc}
    Let $\Theta = \mathcal{X} = \{0,1\}$. Here, the state vector $\vtheta \in \{0,1\}^M$ contains the true costs or qualities of the $M$ items, drawn i.i.d. Each agent $n$ observes, for each item $i$, a noisy signal:
    \[
    \rx_{n,i} = \begin{cases}
        \theta_i & \text{with probability } 2/3\\
        1-\theta_i & \text{otherwise}
    \end{cases}
    \]
\end{example}

The following example illustrates only that SST does not imply SOC.

\begin{example}[RUM with Cauchy Noise]
    Consider $\vtheta,\vx\in \mathbb{R}^2$ in a RUM with Cauchy noise $\epsilon \sim \text{Cauchy}(0,1)$, where the PDF is $f(\epsilon) \propto \frac{1}{1+\epsilon^2}$.
    We compare the density of the signal $\vx$ (which is consistent with $\vtheta$ since $x_1 < x_2$ and $\theta_1 < \theta_2$) against the swapped signal $\vx' = (-4, -5)$:
    \[
        \frac{f_{\rvx|\vtheta}((-5,-4)\mid (0,2))}{f_{\rvx|\vtheta}((-4,-5)\mid (0,2))} = \frac{(1+(-4)^2)(1+(-7)^2)}{(1+(-5)^2)(1+(-6)^2)} = \frac{(17)(50)}{(26)(37)} = \frac{850}{962} < 1.
    \]
    Since the consistent signal is less likely than the inconsistent one, the model violates SOC. However, the model satisfies SST by~\cite{chen2024carrot}.
\end{example}

\paragraph{Necessity of SOC}\label{app:soc_necessity}
The SOC assumption is not only a technical condition for our proof. To illustrate this, fix an item-neutrality peer-only scoring rule \(u(r,s)\) that does not depend on the underlying information structure, where \(r\) is the agent's report and \(s\) is the peer's report. For a comparison between a signal ranking \(\vx\) and its swapped version \(\sigma(\vx)\), call the rule \emph{correctly oriented} if, when the peer is perfectly informative and reports the order consistent with \(\vx\), the rule weakly prefers reporting \(\vx\) to reporting \(\sigma(\vx)\). We call it \emph{non-degenerate} if this preference is strict.

\begin{proposition}[Necessity of SOC]\label{prop:soc_necessity}
If an SOC swap inequality fails, then there is an elicitation instance in which every fixed non-degenerate correctly oriented peer-only scoring rule fails to make truthful reporting a Bayes--Nash equilibrium.
\end{proposition}

\begin{proof}
Suppose SOC fails for some number of items \(M\), state \(\vtheta\), signal ranking \(\vx\), and transposition \(\sigma=\sigma_{ij}\), with \(\vx\) order-consistent with \(\vtheta\) on \((i,j)\):
\[
    P(\vx\mid \vtheta)<P(\sigma(\vx)\mid \vtheta).
\]
Consider an instance whose prior is supported uniformly on the two states \(\vtheta\) and \(\sigma(\vtheta)\). Let the peer be perfectly informative about the state and report the order induced by that state. By item-neutrality of the signal channel,
\[
    P(\vx\mid \sigma(\vtheta))=P(\sigma(\vx)\mid \vtheta).
\]
Thus, after observing \(\vx\),
\[
    \Pr[\vtheta\mid \vx]<\Pr[\sigma(\vtheta)\mid \vx].
\]
The agent therefore believes that a truthful peer is more likely to report the swapped order \(\sigma(\vtheta)\) than the order \(\vtheta\).

For brevity, identify \(\vtheta\) and \(\sigma(\vtheta)\) with the peer reports induced by those states. Define
\[
    A=u(\vx,\vtheta)=u(\sigma(\vx),\sigma(\vtheta)),
    \qquad
    B=u(\sigma(\vx),\vtheta)=u(\vx,\sigma(\vtheta)),
\]
where the equalities follow from item-neutrality of the scoring rule. Correct orientation and non-degeneracy give \(A>B\). Writing \(\pi=\Pr[\vtheta\mid \vx]\), the expected-payoff gain from reporting \(\vx\) rather than \(\sigma(\vx)\) is
\[
    \pi A+(1-\pi)B-\bigl(\pi B+(1-\pi)A\bigr)
    =(2\pi-1)(A-B)<0.
\]
Thus the agent strictly prefers the swapped report to the truthful report. Truthful reporting is therefore not a Bayes--Nash equilibrium in this instance.
\end{proof}

\paragraph{SOC and Order Independence}\label{app:orderindep}    
The SOC property also connects to \emph{order independence} in probabilistic social choice, which is a relaxed form of independence of irrelevant alternatives~\cite{tversky1972elimination}. Specifically, a choice function specifies the probability of selecting an item from a given set. Formally, let $F(i,A) = \Pr[\text{choose } i \mid A]$. A choice function $F$ is \emph{order independent} if for all disjoint subsets $A,B\subset [M]$ and items $i,j\in A$ and $k\in B$:
$$F(i,A) > F(j,A) \iff F(k,B\cup\{i\}) < F(k,B\cup\{j\}).$$

Given a conditional probability $P:\Theta^M\to \Delta(\mathcal{X}^M)$, we define the induced choice function $F_P$ such that $F_P(i,A) = \Pr[\rx_i < \min_{j \in A \setminus \{i\}} \rx_j\mid \vtheta]$ which outputs the top item based on the noisy signals. We use the convention $\min\varnothing=+\infty$.

\begin{proposition}
    If $P$ is SOC, then the induced choice function $F_P$ is order independent.
\end{proposition}

\begin{proof}
    We know $F_P(i,A)>F_P(j,A)$ if and only if $\theta_i<\theta_j$.  Otherwise, if $\theta_i>\theta_j$, by \cref{def:soc}
    \begin{align*}
        F_P(i,A)-F_P(j,A) = &\Pr[x_i<\min_{k\in A\setminus\{i\}}x_k\mid \vtheta]-\Pr[x_j<\min_{k\in A\setminus\{j\}}x_k\mid \vtheta]\\
        =& \sum_{\vx:x_i<\min_{k\in A\setminus\{i\}}x_k} \Pr[\vx\mid\vtheta]-\sum_{\vx:x_i<\min_{k\in A\setminus\{i\}}x_k}\Pr[\sigma_{i,j}(\vx)\mid\vtheta]\\
        =& \sum_{\vx:x_i<\min_{k\in A\setminus\{i\}}x_k} \left(\Pr[\vx\mid\vtheta]-\Pr[\sigma_{i,j}(\vx)\mid\vtheta]\right)<0
    \end{align*}
    which is a contradiction.  The opposite direction is similar.
    
    Finally, given $\theta_i<\theta_j$, we have
    \begin{align*}
        &F_P(k,B\cup\{i\})-F_P(k,B\cup\{j\})\\
        =& \sum_{\vx:x_k<\min\{\min_{\ell\in B\setminus\{k\}}x_\ell,x_i\}} \Pr[\vx\mid\vtheta]\\
        &-\sum_{\vx:x_k<\min\{\min_{\ell\in B\setminus\{k\}}x_\ell,x_j\}}\Pr[\vx\mid\vtheta].
    \end{align*}
    Terms with $x_k < x_i$ and $x_k < x_j$ cancel out. The remaining terms are those where $x_k$ is smaller than one of $\{x_i, x_j\}$ but larger than the other. Thus, the difference reduces to:
\begin{align*}
     &\sum_{\vx: x_j \le x_k < \min\{\min_{\ell\in B\setminus\{k\}}x_\ell,x_i\}} \Pr[\vx\mid\vtheta]\\
     &\quad-\sum_{\vx: x_i \le x_k < \min\{\min_{\ell\in B\setminus\{k\}}x_\ell,x_j\}} \Pr[\vx\mid\vtheta]<0
\end{align*}
by stochastic order transitivity.
\end{proof}

\section{Additional Experimental Results}\label{app:experiments}

\paragraph{Structured unilateral deviations}\label{app:structured_deviations}
We reuse the payment-distribution protocol of~\cref{sec.ecdf} but replace the single uniform-random deviation with a family of structured deviations. This tests the score against near-truthful and coordination-style reports rather than only against a uniformly random report. In each trial, we draw the truthful profile as in~\cref{sec.ecdf}, or the observed-report profile for SUSHI. For a focal agent $n$, peers keep their truthful reports, or their observed reports for SUSHI, while agent $n$'s report is replaced by one of four deviations:
\begin{enumerate}
    \item \emph{uniform}, a uniformly random permutation, matching the unilateral deviation in~\cref{sec.ecdf};
    \item \emph{partial shuffle}, the truthful or observed ranking perturbed by $\lfloor M/2\rfloor$ random adjacent transpositions;
    \item \emph{constant}, a fixed exogenous reference ranking reported regardless of the private signal, capturing a coordination-style report; and
    \item \emph{reverse}, the exact reversal of the truthful or observed ranking.
\end{enumerate}
Aggregating the focal agent's Kendall tau payment over all focal agents and trials yields one empirical payment distribution per deviation. As before, lower ECDFs correspond to more favorable payment distributions.

\begin{figure}[p]
    \centering
    \begin{subfigure}{0.7\linewidth}
        \centering
        \includegraphics[width=\linewidth]{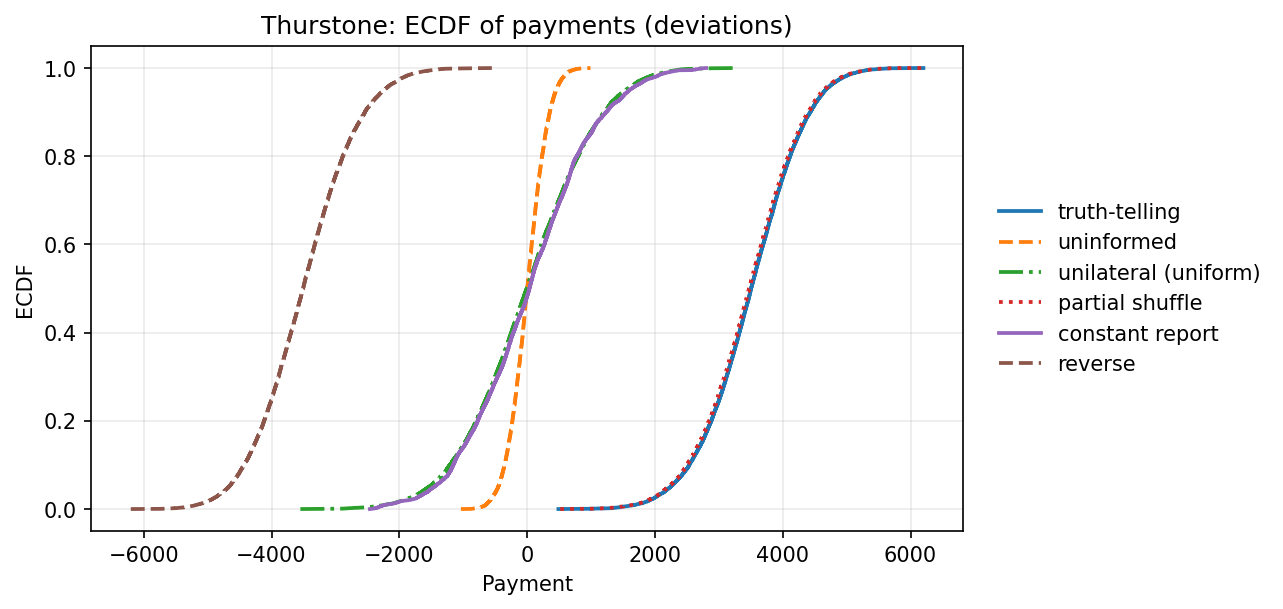}
        \caption{Thurstone}
        \label{fig:dev_thurstone}
    \end{subfigure}

    \begin{subfigure}{0.7\linewidth}
        \centering
        \includegraphics[width=\linewidth]{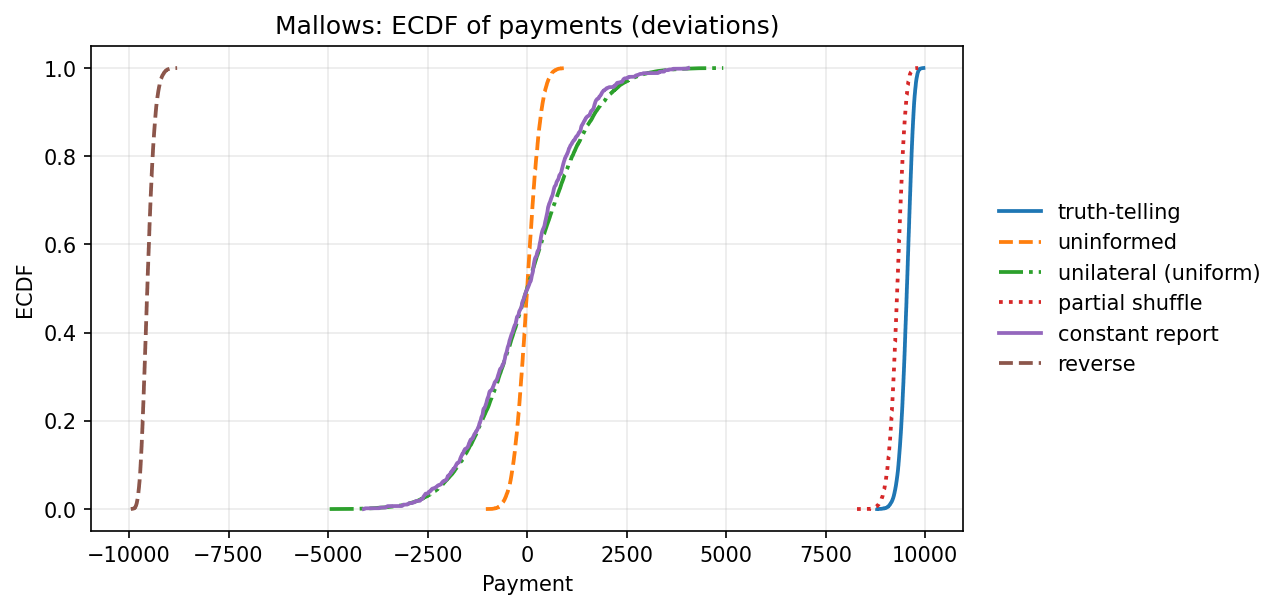}
        \caption{Mallows}
        \label{fig:dev_mallows}
    \end{subfigure}

    \begin{subfigure}{0.7\linewidth}
        \centering
        \includegraphics[width=\linewidth]{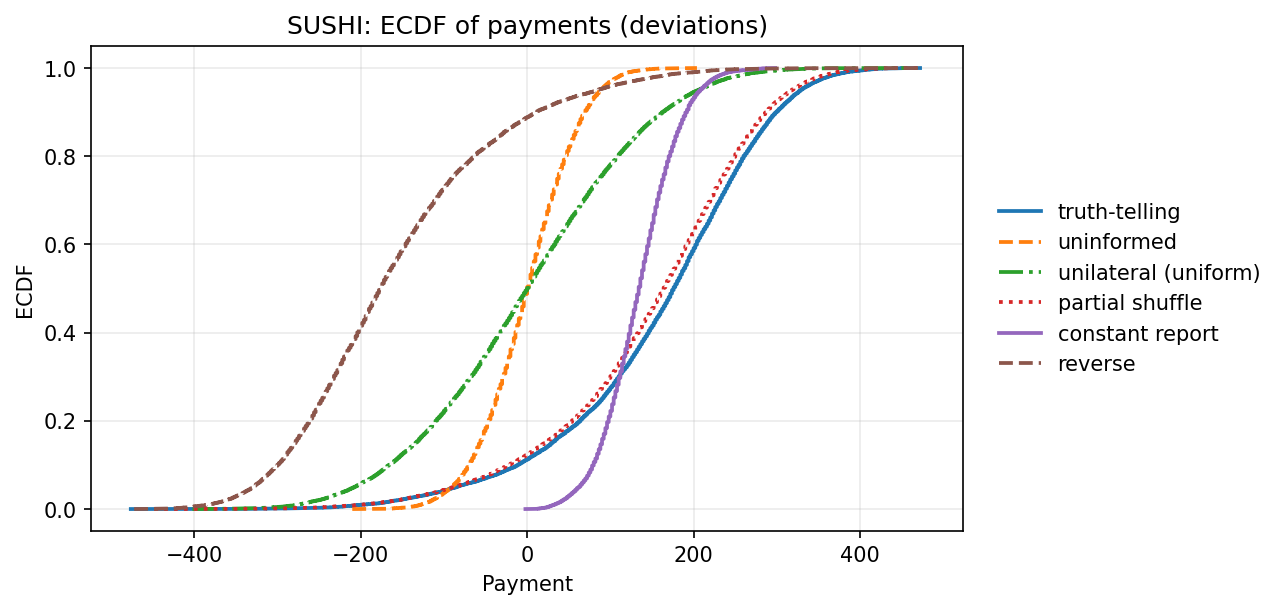}
        \caption{SUSHI}
        \label{fig:dev_sushi}
    \end{subfigure}
    \caption{Structured unilateral deviations. Peers report truthfully, or by observed rankings for SUSHI, while the focal agent reports a uniform random permutation, a partially shuffled ranking, a constant ranking, or the reverse ranking.}
    \label{fig:structured_deviations}
\end{figure}

In the synthetic Thurstone and Mallows models (\cref{fig:dev_thurstone,fig:dev_mallows}), truthful reporting remains right-shifted relative to these structured deviations, with the partial-shuffle deviation closest to truthful reporting and the reverse report performing worst. On SUSHI (\cref{fig:dev_sushi}), the observed-report distribution is also generally the most favorable: although exact pointwise dominance need not hold at every threshold in this real-data stress test, the observed-report ECDF lies below or to the right of the structured-deviation curves over most of the payment range.